\documentclass[3p,11pt]{elsarticle}
\usepackage{lineno,hyperref}
\modulolinenumbers[1]
\usepackage{geometry}
\usepackage{graphicx}
\usepackage[usenames,dvipsnames]{color}
\usepackage{color}
\usepackage[english]{babel}
\usepackage{epstopdf}
\usepackage{mathtools}
\usepackage{latexsym}
\usepackage{hyperref}
\usepackage{url}
\usepackage{amstext}
\usepackage{amssymb}
\usepackage{amsmath}
\usepackage{pifont}
\usepackage{siunitx} 
\usepackage{url}
\hypersetup{
    colorlinks=true,
    linkcolor=blue,
    filecolor=magenta,
    urlcolor=blue,
}
\usepackage{times}
\usepackage{hyperref}
\usepackage{url}
\usepackage{amstext}
\usepackage{amssymb}
\usepackage{wrapfig}
\usepackage{float}
\usepackage{tikz}
\usepackage{pgfplots}
\usepackage{caption}
\usepackage{subcaption}
\usepackage[section]{placeins}

\begin{document}

\begin{frontmatter}

\title{Bayesian inference and role of astrocytes in amyloid-beta dynamics with modelling of Alzheimer's disease using clinical data}

\author[inst1]{Hina Shaheen\corref{cor1}}
\ead{shah8322@mylaurier.ca}
\author[inst1]{Roderick Melnik}
\ead{rmelnik@wlu.ca}

\author[inst3]{ The Alzheimer's Disease Neuroimaging Initiative}
\ead{ADNI}

\cortext[cor1]{Hina Shaheen}
\address[inst1]{MS2 Discovery Interdisciplinary Research Institute, Wilfrid Laurier University, Waterloo, Canada}
\address[inst3]{Data used in preparation of this article were generated by the Alzheimer’s Disease
Metabolomics Consortium (ADMC). As such, the investigators within the ADMC provided
data but did not participate in analysis or writing of this report. A complete listing of ADMC
investigators can be found at: \url{https://sites.duke.edu/adnimetab/team/}}

\begin{abstract}
Alzheimer's disease (AD) is a prominent, worldwide, age-related neurodegenerative disease that currently has no systemic treatment. The symptoms of AD indicate a loss of neuronal function in the human brain. During the disease state, the intricate balance of connections between neurons, astrocytes, microglia, and vascular cells that is required for normal brain function may become altered. Strong evidence suggests that permeable amyloid-$\beta$ peptide ($A\beta$) oligomers, astrogliosis and reactive astrocytosis cause neuronal damage in AD. A large amount of $A\beta$ is secreted by astrocytes, which contributes to the total $A\beta$ deposition in the brain. This suggests that astrocytes may also play a role in AD, leading to increased attention to their dynamics and associated mechanisms. Although astrocytes are the most abundant type of brain cells, even small amounts of $A\beta$ production from a single astrocyte might have a significant impact on the health of the entire brain. Importantly, a deeper understanding of enhanced  astrocyte-neuron interaction might be a potential objective for neuroscience researchers in order to create new therapies capable of modifying astrocytes' activity, reducing neurotoxic consequences, and increasing neuroprotective action. Therefore, in the present study, we developed and evaluated novel stochastic models for $A\beta$ growth using ADNI data to predict the effect of astrocytes on AD progression in a clinical trial. In the AD case, accurate prediction is required for a successful clinical treatment plan. Given that AD studies are observational in nature and involve routine patient visits, stochastic models provide a suitable framework for modelling AD. Using the approximate Bayesian computation (ABC) approach, the AD etiology may be modelled as a multi-state disease process. As a result, we use this approach to examine the weak and strong influence of astrocytes at multiple disease progression stages using ADNI data from the baseline to $2$-year visits for AD patients whose ages ranged from $50$ to $90$ years. Based on ADNI data, we discovered that the strong astrocyte effect (i.e., a higher concentration of astrocytes as compared to $A\beta$) could help to lower or clear the growth of $A\beta$, which is a key to slowing down AD progression. Interestingly, we found that astrocytes whose concentration is less than $A\beta$ increase the production of $A\beta$, which leads to enhanced AD. The proposed model improves our understanding of astrocyte contributions to $A\beta$ growth during AD, which may result in more successful therapeutic approaches. Overall, the evidence strongly supports the involvement of astrocytes in the pathogenesis of neurological disorders like AD, and targeting these cells may hold a promising approach as a therapeutic strategy for such conditions.

\end{abstract}

\begin{keyword}
Large-scale brain \sep multiscale modelling \sep Alzheimer's disease \sep astrocytes \sep  $A\beta$ growth \sep stochastic models \sep inverse problems \sep data-driven methods \sep approximate Bayesian computation technique \sep  parameter estimation \sep neuroscience
\end{keyword}

\end{frontmatter}

\section{Introduction}
The human brain is a system with billions of neurons that is incredibly complex and is managed by cognitive processes. Neurodegenerative diseases (NDDs), such as Alzheimer's disease (AD), are defined by a progressive pathological transformation of the brain's biochemical processes and anatomy, which eventually leads to permanent impairment of cognitive functioning  \cite{lorenzi2019probabilistic}. Amyloid plaques, neurofibrillary tangles, neuropil threads, dystrophic neurites, cerebral amyloid angiopathy, and glial responses are examples of “positive" lesions that accumulate in the brains of AD patients \cite{hyman2012national}. Plaques are extracellular deposits of the amyloid-$\beta$ peptide ($A\beta$) that are created when the amyloid precursor protein is sequentially cleaved by the enzymes $\beta$-secretase and $\gamma$-secretase \cite{henstridge2019beyond}. Years before AD symptoms appear, plaque begins to accumulate in the neocortex, then moves on to the hippocampus, diencephalon, striatum, brainstem, and lastly the cerebellum \cite{thal2002phases}. The formation and deposition of the $A\beta$ peptide are largely thought to be responsible for the development of AD \cite{wang2009tau,nelson2016neurovascular,goodman2022recent,bhagwat2018modeling,hampel2021amyloid}.

The temporal evolution and spatio-temporal distributions of $A\beta$ in AD have been broadly characterized ex vivo through neuropathologic studies utilizing $A\beta$ immunostaining \cite{thal2002phases,whittington2018spatiotemporal}. In terms of the temporal evolution of $A\beta$ in AD, studies suggest that the accumulation of $A\beta$ plaques in the brain may occur years or even decades before the onset of clinical symptoms \cite{arendt2001alzheimer,visser2022cerebrospinal,shaheen2021neuron}. Moreover, $A\beta$ has a regular spatiotemporal distribution that is initially limited to a few specific brain regions, for instance, in the cerebral cortex of patients with AD, before spreading further in the disease state \cite{arnold1991topographical,braak1991neuropathological}. Importantly, the main hypothesis that is often used is that $A\beta$ growth starts in a limited number of affected brain areas and then spreads to other brain regions during the course of the disease via mechanisms like prion-like self-propagation or transsynaptic spread \cite{whittington2018spatiotemporal,song2014beta,nath2012spreading}. The in vivo regional $A\beta$ concentration can be assessed in
humans using the ADNI database \url{(https://adni.loni.usc.edu/)}. There have been large-cohort cross-sectional studies on ADNI for $A\beta$ growth, but the research in this context has mostly concentrated on identifying the different clinical stages of the disease \cite{rowe2010amyloid,mueller2005ways}. The amount of knowledge on the evolution of $A\beta$ plaques over time is still quite limited, despite recent approvals by the FDA of two new drugs in 2021 (aducanumab) and 2023 (lecanemab). In particular, longitudinal analysis has been limited to small time frames of no more than two years or the average deposition throughout the whole brain \cite{resnick2010longitudinal,landau2015measurement,villain2012regional,bateman2012clinical,hansson2018csf}. 
Therefore, as such, the spatiotemporal distributions of $A\beta$ have not been completely defined yet despite the huge number of in vivo data that have been gathered and analyzed.

It is known that abnormal $A\beta$ aggregation, astrogliosis, and reactive astrocytosis are major pathogenic hallmarks of AD \cite{nedergaard2003new,ricci2009astrocyte}. Interestingly, astrocytes are essential for the glymphatic system and the blood-brain barrier and play a significant part in synapse and the neuronal function \cite{ricci2009astrocyte,reid2020astrocytes}. Astrocytes were thought of as brain-supporting cells that provided nutrients and metabolic support for neurons. However, because of their ability to sense changes in neuronal activity by releasing gliotransmitters and gliomodulators and to control the availability of glutamate, GABA, and energy substrates, it is now well known that astrocytes play a crucial role in neuronal function \cite{parpura1994glutamate,volterra2005astrocytes,choi2014human,brandebura2022astrocyte}. As a result, astrocytes are now understood to play a critical role in long-term potentiation, neuronal circuit preservation, and synaptic transmission \cite{reid2020astrocytes,santello2019astrocyte,lee2022function}. Neurotoxic astrocytes and pro-inflammatory disease-associated microglia are spatially and biologically related to $A\beta$ pathology. Indeed, previous research found that reactive astrocytes in AD cluster to areas of degeneration or $A\beta$ plaques \cite{liddelow2017neurotoxic,shi2017apoe4,sadick2019don}. Furthermore, astrocytes contain detectable levels of APP, $\beta$-secretase, and $\gamma$-secretase, and the formulation of these proteins increases in reactive astrocytes under disease conditions such as exposure to proinflammatory cytokines or $A\beta$ oligomers and fibrils in astrocyte cultures from transgenic mice and human AD patient samples \cite{sadick2019don,hong2003interferon,zhao2011contribution}. A statistically significant relationship exists between $A\beta$, microglia, astrocytes, and overall neural function in a special 3D culture system \cite{park20183d}. Finding out how astrocyte manipulation affects $A\beta$ deposition and, eventually, neuronal survival and activity will shed more light on the pathogens of AD. Consequently, the current study has evaluated how astrocytes affect $A\beta$ growth. Although neurons are frequently believed to be the only source of $A\beta$ in AD, there is strong evidence that astrocytes also play a role in the growth of $A\beta$ \cite{hampel2021amyloid,leblanc1997processing}. In particular, when stimulated by a variety of cellular stimuli, astrocytes upregulate the mechanism required for the production of $A\beta$ \cite{frost2017role,batarseh2016amyloid}.

Note that, two competing models have been proposed to explain the hallmarks of AD: the amyloid hypothesis (the neuron-centric model) and the Inverse Warburg hypothesis (the neuron-astrocytic model) \cite{hardy2002amyloid, hardy2006alzheimer,de2016cellular,demetrius2015alzheimer}. According to the neuron-centric model, a variation in the nuclear genome causes an overproduction of $A\beta$, which is harmful to neurons. On the other hand, following the neuron-astrocytic model, the progression of AD is triggered by flaws in the normal energy transduction mechanism, which are caused by mitochondrial dysregulation. Although certain clinical studies of metabolic interventions have yielded promising findings for improving cognitive function, clinical trials of disease-modifying therapies have continuously failed \cite{winblad2016defeating,kivipelto2017rare}. The inadequately understood nature of AD etiology and progression, as well as insufficient trial design, are some of the causes of significant trial failure. Mathematical modelling of AD progression and the development of clinical trial simulations are critical tools for investigating the reasons why clinical trials fail and improving the clinical trial design. The poorly known nature of AD etiology and development restricts the ability to construct robust mechanistic models for reliable prediction of disease progression. There are also semi-mechanistic mathematical models based on inverse problems that have been developed to reflect changes in cognition over time, as measured by errors on various cognitive tests used to assess patients' cognitive abilities, such as the Modified Mini-Mental State Examination (MMSE) and the AD Assessment Scale \cite{ashford2001modeling,rogers2012combining,samtani2012improved,ito2011disease,mar2015fitting,hadjichrysanthou2018development,veitch2022using,hao2022optimal}. The majority of models created to date in the above context are stochastic in nature \cite{lorenzi2019probabilistic,green2011model}. Such models have the important advantage of allowing for variation in model parameters and disease biomarkers for predicting disease progression. 

Inspired by this fact and using ADNI data, we developed and tested a simple novel stochastic model to predict the influence of astrocytes on AD progression in a clinical trial. To describe the dynamics of $A\beta$ growth, we define a stochastic generalization of a logistic growth model. Since AD studies are observational in nature, with patients' health assessed at regular visits, the pathology can be described as a multi-state disease process using the approximate Bayesian computation (ABC) technique. Importantly, ABC is a data-driven technique that uses many low-cost numerical simulations.
ABC also estimates unknown physical or model parameters, as well as their uncertainties, given reference data from real-world experiments or higher-fidelity numerical simulations \cite{beaumont2019approximate}. Using the ABC technique, previous studies have estimated the probability of varying AD severity stages and examined potential factors that may have an impact on these probabilities \cite{hadjichrysanthou2018development,sathiyamoorthi2021deep}. Therefore, we apply this framework to assess the weak and strong impacts of astrocytes at various disease progression stages utilizing ADNI data for AD patients per 2-year visits with respect to their ages. We validated the developed framework using simulated data before applying it to real longitudinal $A\beta$ growth data from ADNI for AD patients by analyzing the corresponding inverse problem. Since the inverse problems are frequently ill-posed because the data may not be sufficient to identify the cause definitively, there may not be an exact solution. What's more, is that determining the cause without additional information is frequently very sensitive to noise level and modelling errors. By representing the unknown as a random variable to emphasize the uncertainty regarding its value, the Bayesian technique offers a flexible and natural method of introducing additional information to augment the noisy data \cite{calvetti2018inverse}. On the other hand, Bayesian inference techniques are crucial in inverse problems because they help us quantify the degree of uncertainty in the solution and provide prior knowledge, which naturally normalizes the data. To emphasize the uncertainty in their estimation, they treat parameters as random variables \cite{xia2022bayesian,konishi2008bayesian}. Therefore, in the present study, the inverse problem can be reformulated as a problem of Bayesian inference. Moreover, to evaluate the uncertainty quantification of stochastic models, we employed differential equations in a probabilistic framework using the moment-closure approximation approach \cite{frohlich2016inference}. Using the ABC technique, we were able to fit the ADNI data for $A\beta$ with respect to the ages of AD patients of our developed models. Our findings revealed that a strong astrocyte effect, where the initial concentration of $A\beta$ is lower than the concentration of astrocytes, can aid in clearing the growth of $A\beta$. Conversely, weak astrocyte effects can contribute to an increase in the production of $A\beta$. These significant results could assist researchers in identifying techniques to slow down the progression of AD.  Our findings shed light on the preclinical phases of AD and reveal the astrocyte's effect on AD progression. Altogether, these findings support the hypothesis that astrocytes may be essential to developing higher cognitive skills in the human brain. Thus, changes in astrocyte physiological functions contribute to brain disease. Furthermore, Bayesian formalism supplements the standard inverse problem paradigm for understanding the pathogenesis of AD.

The rest of the paper is organized as follows: In Section \ref{meth}, we describe our model in terms of its different components: (i) a deterministic growth model of $A\beta$; (ii) a stochastic modelling approach incorporating the strong and weak astrocyte effect and (iii) the ABC technique and uncertainty estimates. In Section \ref{exp}, we set up the experimental data and model the participant dynamics. We present our results and computational simulations based on the developed stochastic models in Section \ref{res}. The computational results have been obtained with an in-house developed MATLAB code, and all data analysis has been carried out in Python. Finally, we discuss our results, conclude our findings, and outline future directions in Sections \ref{dis} and \ref{con}.

\section{Methodology}\label{meth}
It has been found that astrocytes, which are abundantly present in the brain, can play a vital role in slowing down the progression of AD. Studies have suggested that a strong astrocyte effect can help clear the growth of $A\beta$ while weak astrocyte effects can contribute to an increase in its production \cite{miller2018astrocyte,vincent2010astrocytes,sofroniew2010astrocytes}. The present Section highlights the deterministic and statistical modelling approaches incorporating the strong and weak astrocyte effects on $A\beta$ growth via a model applied within the Bayesian setting. This is achieved by defining the stochastic growth model of $A\beta$ to account for individual correlated time series using ADNI data. Additionally, we simulate the growth trajectories of the stochastic model of $A\beta$ using likelihood estimation. This is followed by the parameter estimation and uncertainty quantification presented in Sub-section \ref{abc}. 

\subsection{Deterministic growth model of $A\beta$ incorporating strong and weak astrocyte effects}\label{sectionm}
\begin{figure}[h]
    \centering
    \includegraphics[scale=0.5]{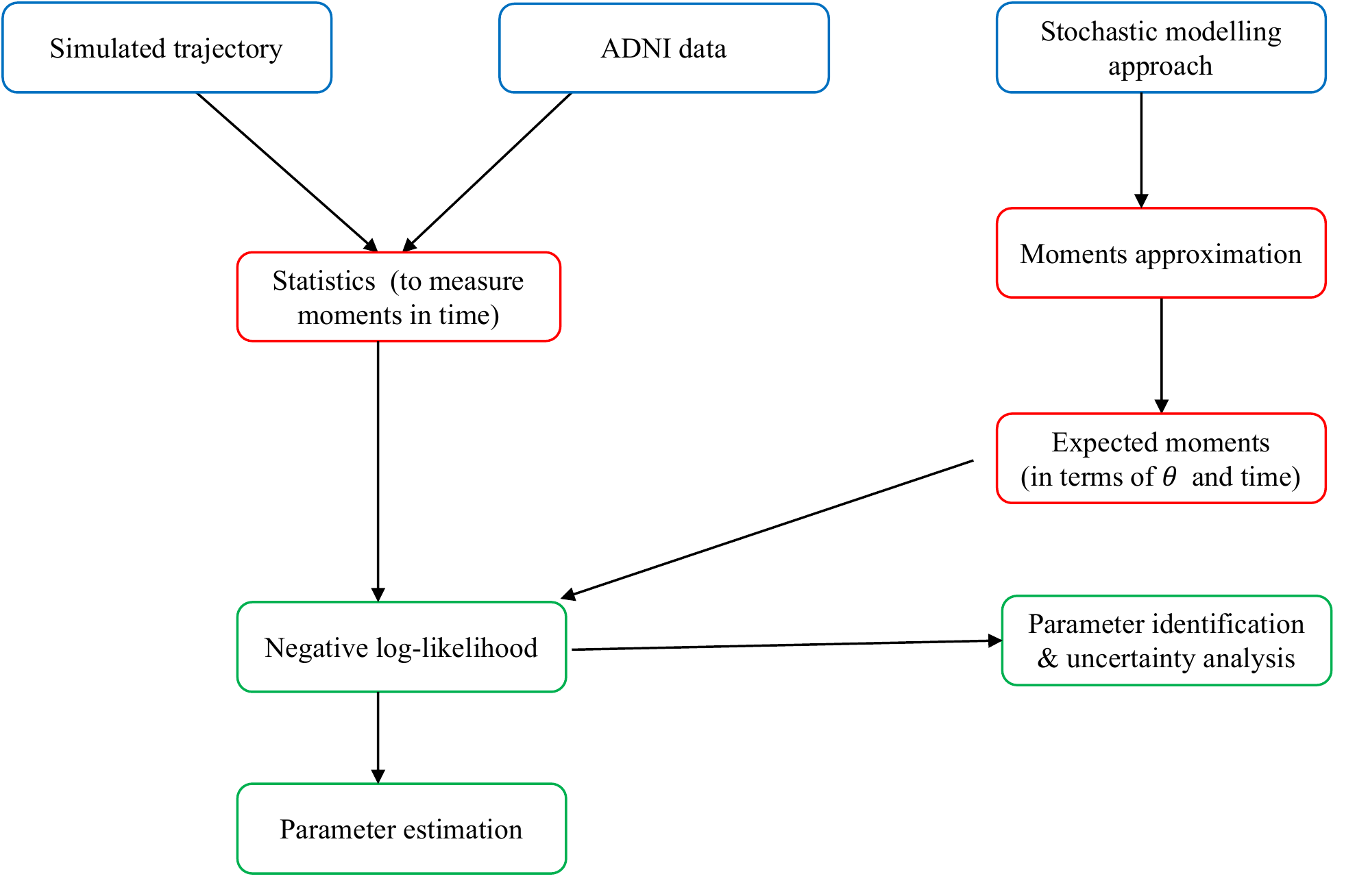}
    \caption{(Color online) Schematic representation of the moment closure approximation (MC) technique for stochastic parameter estimation for the present study. This diagram represents also the structure for the MC technique to obtain moments of a stochastic process and demonstrate how model expected moments are fitted to stochastic data (motivated by \cite{johnson2019cancer}). }
    \label{fig:1}
\end{figure}

\begin{figure}[h] 
\centering
\begin{subfigure}{0.32\textwidth}
\includegraphics[width=\linewidth]{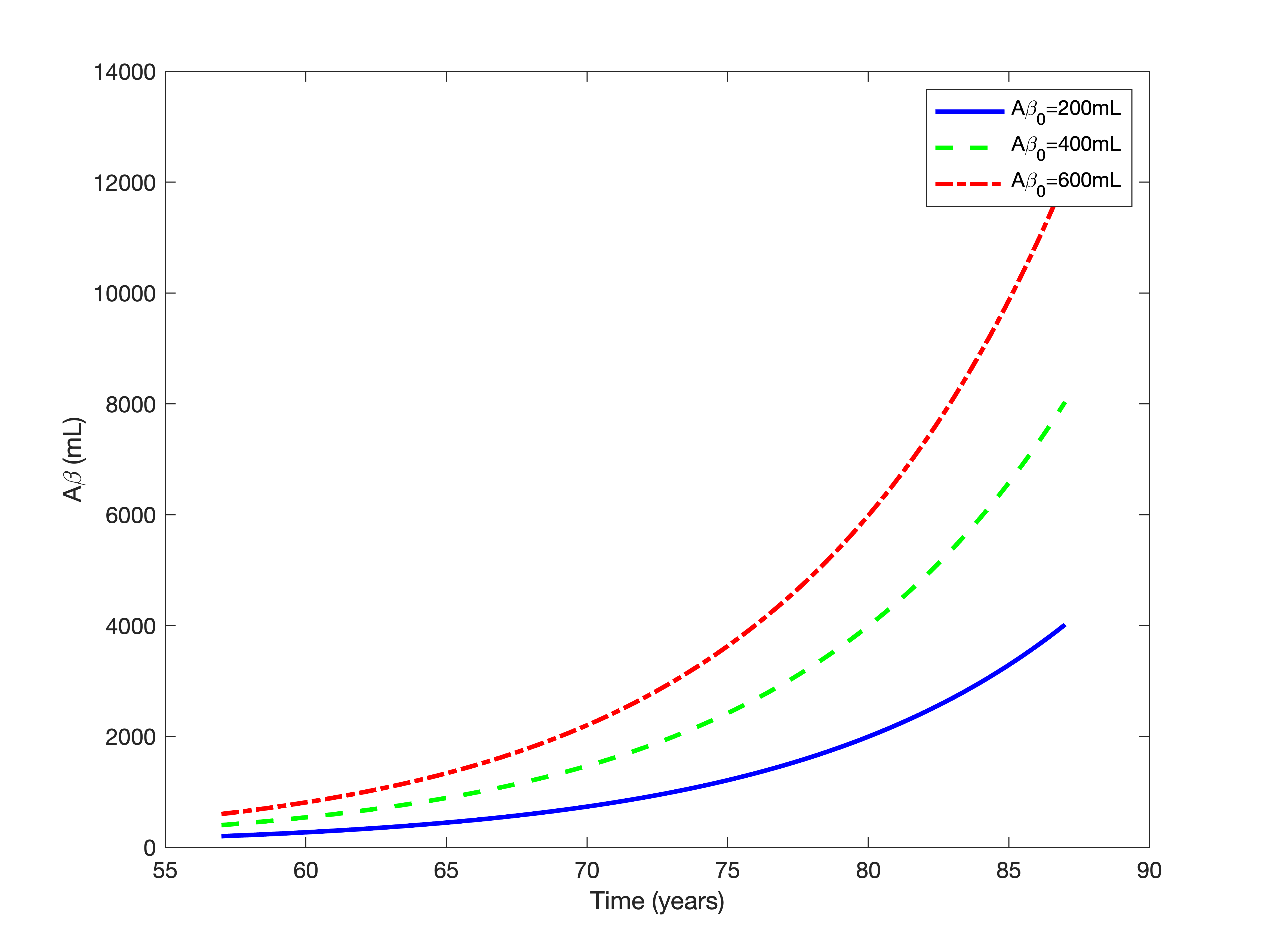}
\caption{} \label{fig:a}
\end{subfigure}
\begin{subfigure}{0.32\textwidth}
\includegraphics[width=\linewidth]{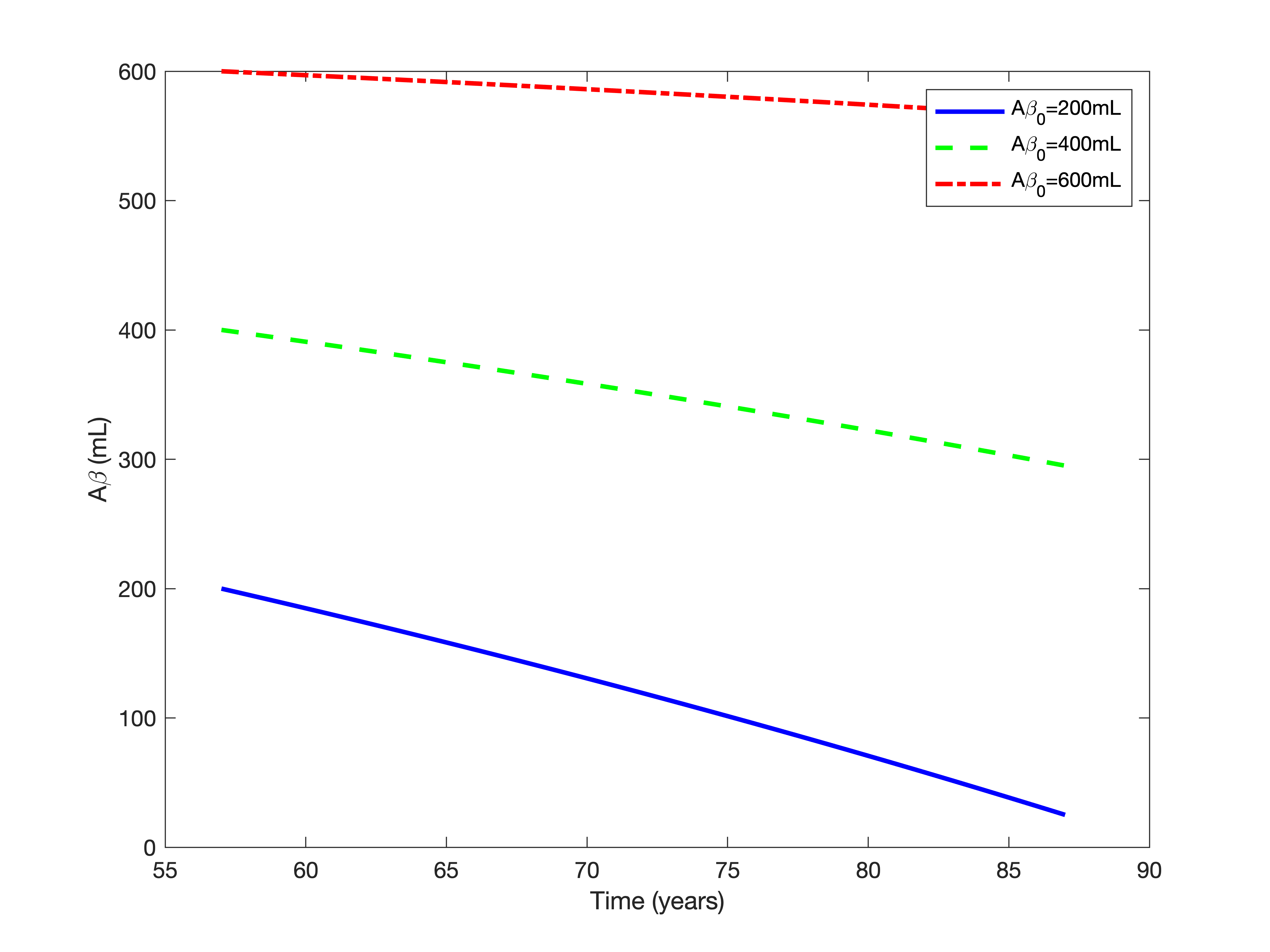}
\caption{} \label{fig:b}
\end{subfigure}
\begin{subfigure}{0.32\textwidth}
\includegraphics[width=\linewidth]{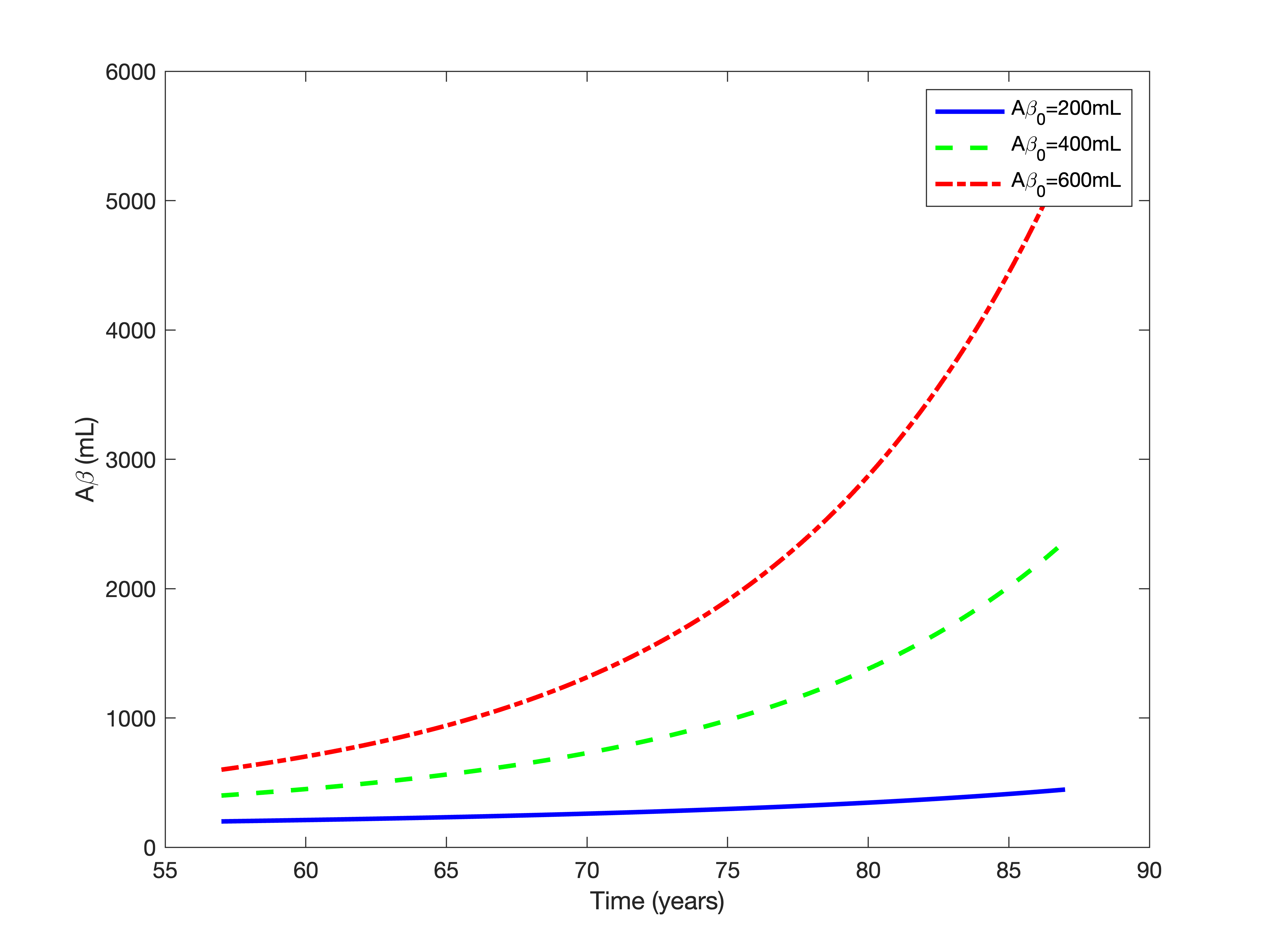}
\caption{} \label{fig:c}
\end{subfigure}
   \caption{(Color online) The dynamic curves for deterministic models for $A\beta_0=200\si{mL}$ (blue), $400\si{mL}$ (green), $600\si{mL}$ (red) (a-c). Here: (a) is the behaviour of $A\beta$ growth model (Eq. \ref{eq1}), (b) is the strong astrocyte effect on  $A\beta$ growth model (Eq. \ref{astro}), and (c) is the weak astrocyte effect on  $A\beta$ growth model (Eq. \ref{astro1}).}
    \label{fig:2}
\end{figure}

Fig. \ref{fig:1} gives an overview of our methods. As can be seen, our end goal is to focus on the stochastic growth model of $A\beta$ because of the inherent stochasticity of monomers growth processes in the regime of small $A\beta$ peptides numbers \cite{meisl2022mechanistic}. However, the structure of the functional forms of the kinetic equations describing the deposition of $A\beta$ proteins in the brain can be understood fairly well within a deterministic framework. Therefore, this step will link us to the historical and most widely implemented mathematical models of AD in neuroscience. In the present study, we consider the following three deterministic models (given in Eqs. (\ref{eq1}), (\ref{astro}-\ref{astro1})) that describe the kinetic growth of $A\beta$. The first model is the baseline model of $A\beta$ growth, in which the growth rate $ \frac{d A\beta(t)}{d t}$ is proportional to the number of $A\beta$, and a single growth rate constant, $r$, resulting in the standard exponential growth model:
\begin{equation}\label{eq1}
    \frac{d A\beta(t)}{d t}=rA\beta(t),
\end{equation}
where $A\beta(t)=A\beta(0)e^{rt}$ and $A\beta(0)$ is the initial value of $A\beta$ at time $t$.
This model represents the evolution of $A\beta$ monomers that display individual autonomous proliferation provided by the growth rate constant $r$ across time and $A\beta$ as shown in Fig. \ref{fig:2} (a) for initial values of $A\beta$ equals $[200; 400; 600]mL$. We choose these values for $A\beta(0)$ so that they would be consistent with the ADNI data for AD patients who had baseline (bl) to $2$-year visits (i.e., $A\beta(0)= 200mL$ at baseline, $A\beta_{\rm 12-month-visits} (57)= 400mL$ at $12$ months, and $A\beta_{\rm 24-month-visits} (57)= 600mL$ at $24$ months are the initial values for the mean of $A\beta$ at three visits). Moreover, the results are plotted as a function of age in years, assuming that the participants are all $57$ years old at baseline, as seen in Fig. \ref{fig:2} (a). Since AD is a disease that typically affects individuals over the age of $50$, however, It's also possible that the modelling approach used in the  present study could be applied to larger datasets or to different age groups, which could help to better understand the dynamics of $A\beta$ accumulation and progression of AD.

Initially, AD targets neurons and their connections in brain regions associated with memory, such as the entorhinal cortex and hippocampus. Subsequently, it progresses to affect other areas in the cerebral cortex responsible for functions such as language, reasoning, and social behaviour \cite{grieco2023probing}. Notably, the logistic growth model has been used to model the growth of $A\beta$ concentration ($mL$) over time \cite{whittington2018spatiotemporal,lee2007three}. $A\beta$ is initially generated in a limited number of affected brain regions and, over the course of the disease, can propagate to other regions of the brain. Thus, in the classical formulation of the logistic growth model for $A\beta$, the growth rate is characterized by a growth rate constant ($r$) modulated by an additional term for the temporal evolution to describe the slowing of the growth rate as the population approaches carrying capacity ($g$): 
\begin{equation}\label{abeta}
    \frac{d A\beta(t)}{d t}=r(1-\frac{A\beta(t)}{g})A\beta(t).
\end{equation}
Importantly, in the present study, we are considering the temporal evolution of the growth of $A\beta$. The logistic growth models that describe two temporal regimes of brain involvement in affected areas during AD: when $A\beta<<g$, the $\frac{A\beta}{g}$ is negligible and the $A\beta$ plaques essentially exhibit exponential growth, and when $A\beta$ is near $g$, the net growth rate $\frac{d A\beta}{d t}$ slows towards zero, as $A\beta$ approaches $g$ and the term $(1-\frac{A\beta}{g})$ approaches zero. 

Moreover, when astroglia is not overactivated in the early stages of AD, they destroy the soluble form of $A\beta$ through apolipoproteins, $ApoE2$ \cite{kim2018exploring}. But if astroglia is moderately activated, they protect neurons and produce a drop in the growth of $A\beta$ \cite{thuraisingham2022kinetic}.  However, as AD progresses, astrocytes feed on themselves, resulting in the overactivation of astroglia. Furthermore, astrocytes are said to boost $A\beta$ deposition by overexpressing $\beta$-secretase BACE1, which cleaves the APP alongside $\gamma$-secretase \cite{ricci2009astrocyte}. Astrocytes are usually triggered by extracellular TNF-$\alpha$, but they are also stimulated by $A\beta$ \cite{hao2016mathematical}. Indeed, the precise involvement of astrocytes in $A\beta$ clearance and/or deposition is still unknown \cite{heneka2007inflammatory,rocchi2003causative}. Research has shown that astrocytes can have both strong and weak effects on the growth of $A\beta$, depending on the specific type of astrocyte \cite{miller2018astrocyte,vincent2010astrocytes,sofroniew2010astrocytes}. Strong astrocytes refer to astrocytes that have a more activated state and are more responsive to changes in their environment. These astrocytes have been shown to have a protective effect on neurons and can help to reduce the growth of $A\beta$. Strong astrocytes can do this by releasing factors that promote the clearance of $A\beta$, such as apolipoprotein E, or by secreting factors that promote the growth of new neurons, which can help to offset the damage caused by $A\beta$. On the other hand, weak astrocytes refer to astrocytes that are less activated and less responsive to changes in their environment. These astrocytes have been shown to have a weaker protective effect on neurons and can actually promote the growth of $A\beta$. Weak astrocytes can do this by releasing factors that promote the aggregation of $A\beta$ or by failing to clear $A\beta$ from the brain. As a result, we provide the logistic growth model to highlight the strong influence of astrocytes, which is similar to the logistic growth model except that the reliance on $A\beta$ happens in the opposite regime in this model \cite{miller2018astrocyte,vincent2010astrocytes,sofroniew2010astrocytes,deane2008apoe,palabas2023double}. We introduced here the term of astrocyte effect in the form of $1-\frac{A_{astro}}{A\beta}$ that lowers the observed growth rate at small $A\beta$ near the threshold $A_{astro}$ as:
\begin{equation}\label{astro}
    \frac{d A\beta(t)}{d t}=r(1-\frac{A_{astro}}{A\beta(t)})A\beta(t).
\end{equation}
If the concentration of $A\beta$ is lower than the concentration of astrocytes, then astrocytes can have a strong effect on $A\beta$ metabolism and clearance. Therefore, we choose $A_{astro}=700\si{mL}$ which results in the clearance of $A\beta$ as shown in Fig. \ref{fig:2} (c). As expected, initial conditions lower than the concentration of $A_{astro}=700\si{mL}$ (corresponding to $A\beta_0=200$, $A\beta_0=400$ and $A\beta_0=600$) result in a slowing down of $A\beta$ net growth rate. However, it is important to note that this clearance rate may change over time due to various factors that can influence the system. The value of $A_{astro}$ is chosen based on the ideas highlighted in \cite{miller2018astrocyte,thuraisingham2022kinetic,carter2019astrocyte}. Moreover, the fact that a decrease in the concentration of astrocytes i.e., AD may start to propagate once we decrease the $A_{astro}$ would lead to an increase in the concentration of $A\beta$ \cite{pal2022influence,pal2022coupled}. This model is able to explain the threshold-like behaviour of astrocytes observed in preclinical studies of the progression of AD in mice \cite{carter2019astrocyte,olsen2018astroglial,lesne2003transforming}, where below a threshold number of affected brain areas, AD never forms. 
Finally, in order to incorporate weak astrocyte effects, where we consider the growth rate to be always greater than zero for the initial value of $A\beta$, we introduce an additional model. We call it the extended astrocyte deterministic model (i.e., the weak astrocyte effect model), and it is given as follows \cite{batarseh2016amyloid,palabas2023double,blasko2000costimulatory}:
\begin{equation}\label{astro1}
    \frac{d A\beta(t)}{d t}=r(1-\frac{A_{astro}+\gamma}{A\beta(t)+\gamma})A\beta(t),
\end{equation}
where $\gamma$ presents the clearance rate adopted from \cite{fornari2020spatially}. This model is similar to the strong astrocyte effect model (Eq. \ref{astro}), but it includes a new parameter $\gamma$ that allows the model to show down a strong astrocyte impact when $A_ {astro}>A\beta$. However, for weak astrocyte conditions, the term $A_ {astro}$ remains less than $A\beta$, maintaining a net growth rate as $A\beta$ increases. The behaviour of the extended astrocyte model with lower concentrations of $A_{astro}$ that result in a weak astrocyte effect is shown in Fig. \ref{fig:2} (c). Potentially weak astrocyte effects, such as seen during a metabolism reaction, are explained by the weak astrocyte deterministic model \cite{sadick2019don}. The three aforementioned growth models based on Eq. (\ref{eq1}) and Eqs. (\ref{astro}-\ref{astro1}) have been used for obtaining the results presented in Fig. \ref{fig:2}. The parameter $r= 0.0227$, is the same for all Eqs. (\ref{eq1}, \ref{astro}-\ref{astro1}) (adopted from \cite{fornari2020spatially}), however, for Eq. (\ref{astro}), $A_{astro} =700\si{mL}$ and for Eq. (\ref{astro1}), $A_{astro}= 100\si{mL}$. The values of $A_{astro}$ are chosen based on the facts  that a reduction in the concentration of astrocytes (i.e., in AD) may trigger the spread of the disease suggesting that a decrease in $A_{astro}$ could lead to an increase in the concentration of $A\beta$ \cite{whittington2018spatiotemporal,thuraisingham2022kinetic,carter2019astrocyte}. It is well known that, a logistic growth model, in which the carrying capacity of $A\beta$ varies across the brain but the exponential growth rate and duration of half the maximum $A\beta$ concentration are constant, can be used to mathematically simulate the in vivo temporal evolution of $A\beta$ in AD \cite{hampel2021amyloid,bertens2015temporal,jagust2021temporal}.

\subsection{The stochastic modelling approach} \label{stoch}
Here, we make the assumption that the $A\beta$ growth kinetics are observed in a subset of the affected areas of the brain, where neural activity is expected to vary largely in response to ``birth" and ``death" events, which together define the net growth rate, $\frac{d A\beta}{d t}$. This situation results in apparent growth kinetics of $A\beta$, which differ markedly from the typical behaviour described in Section \ref{sectionm}. Here, the birth rate is referred to as ``aggregation" and the death rate as ``fragmentation". This development follows the ideas of \cite{fornari2020spatially}. A simple model (i.e., Eq. (\ref{eq1})) for the formation of monomers includes linear aggregation ($r_a$), which represents the specific scenario in which monomers are added to an aggregate, and fragmentation ($r_f$), which represents the specific scenario in which monomers are created \cite{fornari2020spatially}. A stochastic modelling framework was used to evaluate the applicability of the aforementioned astrocyte effect models (i.e Eqs. (\ref{astro}-\ref{astro1})). We selected the formulations for our stochastic models in this framework to recapitulate a first-order moment (mean) that matched the equivalent deterministic ODE of the $A\beta$ growth model and strong and weak astrocyte effect on $A\beta$ growth presented in Eqs. (\ref{eq1}, \ref{astro}-\ref{astro1}). Thus, fragmentation and aggregation rates were chosen based on plausible predictions that were consistent with the astrocyte effect model behaviour. The majority of the models developed so far for AD progression have a stochastic aspect \cite{hadjichrysanthou2018development,bilgel2019predicting,burnham2020impact}. Such models allow for variation in the model's parameters and stochastic factors in the development of AD pathology. In the context of the stochastic model, $x(t)$ refers to the random variable that represents the concentration of $A\beta$ at time $t$. This random variable takes on different values in different trials of the experiment due to the inherent randomness in the system. The expected value of $x(t)$, denoted as $\langle x(t)\rangle$, represents the average concentration of $A\beta$ that we expect to observe at time $t$ if we repeated the experiment many times.

The goal of the stochastic model is to capture the randomness in the system and predict the probability distribution of $x(t)$ at different times. In developing such a model in our context, we follow the ideas of \cite{johnson2019cancer}. The expected mean number for $A\beta$ concentration in time $\langle x(t)\rangle$ is identical to what the deterministic models anticipate, which means that the average value of $x(t)$ predicted by the stochastic model is the same as the solution of the deterministic models given by Eqs. (\ref{eq1}, \ref{astro}-\ref{astro1}). Therefore, for the stochastic models, the time evolution of $x(t)$ in a simple way is defined as:
\begin{itemize}
  \item Event 1: aggregation $C+C \rightarrow 2C$ (reaction rate: $r_a(x)$)
  \item Event 2: fragmentation $2C \rightarrow C+C$ (reaction rate: $r_f(x)$),
\end{itemize}
where $C$ denotes monomers \cite{fornari2020spatially}, $r_a(x)$ and $r_f(x)$ describe the rates at which the corresponding event occurs. Given a stochastic process with a set of possible events $i$, generally, the probability of event $i$ taking place during an infinitesimal time interval $\Delta t$ can be expressed as the multiplication of the event rate, the population state, and the duration of the time interval given as \cite{johnson2019cancer}:
\begin{equation} \label{p1}
    P_{event}=r_{event}(x)x\Delta t.
\end{equation}
The chance of an event occurring is always a function of $x$ since it is a first-order reaction, and we restrict the potential events to aggregation or fragmentation events \cite{fornari2020spatially}. The aggregation rate ($r_a$) and fragmentation rate ($r_f$) in the basic aggregation-fragmentation model are independent of $x$ (i.e., $r_a=r_f=r$). For instance, for a simple growth model (i.e Eq. \ref{eq1}), 
\begin{equation}\label{s1}
    p_{a}=rx\Delta t,\,\,\ p_{f}=rx\Delta t.
\end{equation}
The above model's typical behaviour matches the exponential growth model. However, we incorporate aggregation and/or fragmentation rates that are functions of $x$ for the remaining stochastic model. In the stochastic model that we consider, the aggregation and/or fragmentation rates are not assumed to be constant but rather depend on the size of the monomer or aggregate. Specifically, we incorporate aggregation and/or fragmentation rates that are functions of $x$, denoted as $r_a(x)$ and $r_a(x)$, respectively. These rates capture the fact that the probability of aggregation or fragmentation occurring at a given time depends on the size of the monomer or aggregate and the likelihood of interactions between them \cite{liu2020stochastic}. The first stochastic astrocyte model we consider is the strong astrocyte effect model ($sM1)$, presented in Table \ref{tab:1}. In this model, we assume that for $x$ approaching the threshold value $A_{astro}$, the astrocyte effect has an equal effect on reducing birth probabilities ($p_a$) and increasing death probabilities ($p_f$), leading to: 
\begin{equation}\label{ast1}
    p_{a}=(r_a-\frac{A_{astro}}{x}\frac{(r_a-r_f)}{2})x\Delta t,
\,\,\
    p_{f}=(r_f+\frac{A_{astro}}{x}\frac{(r_a-r_f)}{2})x\Delta t.
\end{equation}
The weak astrocyte effect model ($sM2$) presented in Table \ref{tab:1} may be demonstrated using the same justifications as the strong astrocyte effect model, resulting in the following birth and death probabilities:
\begin{equation}\label{ast3}
    p_{a}=(r_a-\frac{A_{astro}+\gamma}{x+\gamma}\frac{(r_a-r_f)}{2})x\Delta t,
\,\,\
    p_{f}=(r_f+\frac{A_{astro}+\gamma}{x+\gamma}\frac{(r_a-r_f)}{2})x\Delta t,
\end{equation}
where $\gamma$ is the clearance rate adopted from \cite{fornari2020spatially}. The models presented in Table \ref{tab:1} are stochastic models and we treat them as an inverse problem. Each of these models has a set of parameters that determine the probability of observing a certain event during a time interval. In Table \ref{tab:1}, each of the three stochastic models with their probabilities mentioned above i.e., Eqs. (\ref{s1}-\ref{ast3}) are described, along with the assumptions made on the mechanism of aggregation or fragmentation rates.
\begin{table}
\caption{Families of stochastic growth models whose typical behaviour is consistent with deterministic growth models (presented in Section \ref{sectionm}). For the astrocyte model families, the astrocyte effect can alter the probability of aggregation (birth), fragmentation (death), or both, corresponding to various structural interpretations.}
\label{tab:1}       
\fontsize{8pt}{8pt}
%
%
\begin{tabular}{ |p{4cm}|p{6cm}|p{6cm}|  }
\hline
Growth model ($sM1$)& Strong astrocyte effect model ($sM2$) & Weak astrocyte effect model ($sM3$)\\
\hline
\multicolumn{3}{|c|}{The mean of $A\beta$ concentration change is given as a deterministic ODE for each of the stochastic models} \\
\hline
  $\frac{d A\beta(t)}{d t}=rA\beta(t)$  & $\frac{d A\beta(t)}{d t}=r(1-\frac{A_{astro}}{A\beta(t)})A\beta(t)$& ~$    \frac{d A\beta(t)}{d t}=r(1-\frac{A_{astro}+\gamma}{A\beta(t)+\gamma})A\beta(t)$ \\
  \hline
\multicolumn{3}{|c|}{Event probabilities to describe each stochastic model} \\
\hline
on birth \& death &on birth \& death & on birth \& death  \\
\hline
  $P_{a}=(rA\beta)\Delta t$& $    p_{a}=\big(r_aA\beta-A_{astro}(\frac{r_a-r_f}{2})\big)\Delta t$   &  $ p_{a}=\big(r_aA\beta-A\beta(\frac{r_a-r_f}{2})(\frac{A_{astro}+\gamma}{A\beta+\gamma})\big)\Delta t$\\
  \hline
 $P_{f}=(rA\beta)\Delta t$ & $p_{f}=\big(r_fA\beta+A_{astro}(\frac{r_a-r_f}{2})\big)\Delta t$  &
$ p_{f}=\big(r_fA\beta+A\beta(\frac{r_a-r_f}{2})(\frac{A_{astro}+\gamma}{A\beta+\gamma})\big)\Delta t$ \\

\hline
\end{tabular}
\end{table}
We simulate the growth trajectories of the stochastic models using the maximum likelihood method, which involves generating multiple realizations of the corresponding stochastic process and estimating the parameters that best fit the observed data. The stochastic processes referred to in this context are the aggregation and fragmentation events of $A\beta$. These events are modelled as random events that occur over time and result in changes in the concentration of $A\beta$. Specifically, we start with an initial condition for the concentration of $A\beta$, represented by the random variable $x_0$, and simulate the stochastic process by generating a sequence of random events that correspond to the aggregation and fragmentation events of $A\beta$. We repeat this process multiple times to generate a set of simulated growth trajectories, which are then compared to the observed data to estimate the parameters of the stochastic model. The maximum likelihood method allows us to estimate the most likely values of the model parameters based on the observed data, and to quantify the uncertainty in those estimates. This approach provides a way to assess the goodness of fit of the stochastic model to the data and to make predictions about the future behaviour of the system. The models indicated above are used to assess the significance of the astrocyte effect on the growth of the $A\beta$ concentration in the brain during AD. In the stochastic exponential growth model, i.e., $sM1$ (see Table \ref{tab:1}), the growth rate, which is determined as the sum of the birth and death rates, is assumed to be constant and independent of the initial conditions. We examined the average behaviours of the candidate models in their deterministic forms, which allowed us to develop corresponding stochastic models that produce the same average behaviour. The stochastic models are shown in Table \ref{tab:1}, and they were developed based on the deterministic models in order to capture the inherent variability and randomness observed in the present study. In the context of stochastic models, a moment-closure approximation can be used to derive approximate likelihood functions for Bayesian inference \cite{frohlich2016inference}. Therefore, in the next Section \ref{abc}, applying the moment-closure approximation approach for parameter estimation (Fig. \ref{fig:1}) using stochastic models (i.e., $sM1-sM3$) to the high-throughput $A\beta$ growth data from the ADNI database for AD patients allows us to evaluate the applicability of the suggested stochastic models. By computationally generating growth trajectories using a model of intermediate complexity, we first verified our approach. Its ability to correctly identify the underlying model and the true parameters was demonstrated by applying the statistical inference framework and parameter estimation to the generated data, as can be seen next (Section \ref{abc}).

\subsection{Approximate Bayesian Computation (ABC) technique and uncertainty estimates}\label{abc}

First, we note that the ABC methodology is used here to estimate the parameters of stochastic models ($sM1-sM3$) and quantify their uncertainty  \cite{beaumont2002approximate}. The ABC approach involves simulating synthetic data from the model using a set of candidate parameter values, and then comparing these simulated data to the observed data to determine the parameter values that are most consistent with the observed data. Additionally, moment closure and Bayesian techniques can be combined to estimate the parameters of a stochastic model and quantify their uncertainty \cite{robert2011lack}. We use the moment-closure approximation approach presented in Fröhlich and colleagues' work to fit the developed stochastic growth models (i.e., $sM1-sM3$) to experimentally measured growth curves of $A\beta$ from the ADNI database, making inference on the stochastic process possible \cite{frohlich2016inference}. Here we use the 
master equation (ME) to define the change in the distribution of probability that the system has
 (i.e., in this case, $x$) as a function of time. We can derive the expected values of $x_1,x_2,\ldots, x_m$ (with $m$ being the number of observations), and the time derivatives of moments through ME. We developed stochastic models in this context so that the differential of the first-order moment correlates to one of the deterministic models presented in Eqs. (\ref{eq1},\ref{astro}-\ref{astro1}). The discrete-time version of the ME, also known as the Kolmogorov forward equation or the Chapman-Kolmogorov equation, defines the probability of $x$ at time $t$ as a sum of the probabilities of aggregation and fragmentation, the concentration of $A\beta$ can only change by $1$ unit at a time in this model because we assume that the system can only transition between adjacent states \cite{fornari2019prion}. This means that if the system is currently in state $x$, it can only transition to state $x-1$ (decrease by $1$), state $x$ (remain the same), or state $x+1$ (increase by $1$) at the next time step. This assumption of only allowing transitions between adjacent states is based on the idea that the system is subject to some kind of physical or biological constraint that prevents it from making large jumps in concentration. For example, in a biological system, the production and degradation of $A\beta$ may be regulated by enzymes and other molecular processes that operate at a certain rate, which limits the rate at which the concentration can change. By carefully modelling the system in this way, the discrete-time Chapman-Kolmogorov equation allows us to study the dynamics of the system and make predictions about how it will evolve over time, even in the presence of randomness or uncertainty. It is given as follows:
\begin{equation}\label{PDF}
P(x,t+\Delta t) = P(x-1,t) \cdot p_f \cdot \Delta t + P(x,t) \cdot \left[1 - (p_a + p_f) \cdot \Delta t \right] + P(x+1,t) \cdot p_a \cdot \Delta t.
\end{equation}
Here, $P(x,t)$ represents the probability density function of the concentration of $A\beta$ at time $t$, with $x$ representing the possible values of the concentration. The three terms on the right-hand side of the equation (\ref{PDF}) represent the rates of change in the probability density function due to the occurrence of different events. This master equation is a key tool for analyzing the behaviour of stochastic models and can be used to compute various statistical properties of the system, such as the mean and variance of the concentration of $A\beta(t)$ over time. We hope to be able to estimate the mean and variance of $x$ over time using experimental data with adequate repetitions. We want to use ME to derive the first and second moments to directly compare the mean and variance in the ADNI data to the expected mean and variance of each stochastic model (adopted from \cite{johnson2019cancer}). We followed the procedure outlined in \cite{bronstein2018variational}. The mean and variance of each model can be expressed in terms of the first and second-order moments, denoted by $\langle x \rangle$ and $\langle x \rangle^2$, given as:
\begin{equation}
\langle x \rangle (t+\Delta t) = \sum_{x=0}^{\infty} x P(x,t+\Delta t), \,\
\langle x^2 \rangle (t+\Delta t) = \sum_{x=0}^{\infty} x^2 P(x,t+\Delta t).
\end{equation}

Here, the notation $\langle \ldots \rangle$ indicates the expectation value of the moment for each stochastic model ($sM1-sM3$) as presented in Table \ref{tab:2}. The stochastic models have the same mean ODE corresponding to their deterministic model family (exponential, strong astrocyte effect, and weak astrocyte effect models; Eqs. (\ref{eq1}), (\ref{astro}-\ref{astro1})), but they differ in variance depending on whether the astrocyte effect changes the aggregation, fragmentation, or both event terms. To obtain the mean and variance of each stochastic model, it is necessary to compute the first and second moments of the probability distribution function. This can be done by using the ME of each model and following a procedure outlined in \cite{johnson2019cancer}. The estimated mean and variance of the model were obtained by fitting the stochastic model to experimental or simulated data. Then, we estimated the model parameters using a maximum likelihood estimation. Once the model parameters were estimated, the mean and variance of the stochastic model were calculated using the equations for the mean and variance in terms of moments as given in Table \ref{tab:2}. To validate the model, the simulated data set of $10,000$ trajectories was generated using the same model parameters, initial conditions, and time points as simulated data used to estimate the model parameters such as $r_a,r_f$ and $A_{astro}$. The mean and variance of the $10,000$ trajectories were then calculated and compared to the estimated mean and estimated variance of the model. We found that the mean and variance of the simulated data set of $10,000$ trajectories with known parameters matched the estimated mean and variance of the model. The unknown parameters (i.e. denoted as $\theta$) for each stochastic model ($sM1-sM3$), developed in the present study are $r_a,r_f$ and $A_{astro}$, where for simplicity we set $\gamma$ as a known parameter.
\begin{table}
\caption{Using moment approach (motivated by \cite{johnson2019cancer}): Differential equations of the moment-closure approximations of the mean and variance computed from the ME for each stochastic model i.e., Eqs (\ref{s1}-\ref{ast3}) ($sM1-sM3$).}
\label{tab:2}       
\fontsize{8pt}{8pt}
%
%
\begin{tabular}{p{1cm}p{5cm}p{9cm}}
\hline\noalign{\smallskip}
Model& Mean & Variance \\
\hline\noalign{\smallskip}
~$sM1$  & $\frac{d\mu_i}{d t}=\mu_i (r_a-r_f)  $& ~$\frac{d \sigma_i}{d t} =2\sigma_i(r_a-r_f) + \mu_i(r_a+r_f)$ \\
\\
~$sM2$   & $\frac{d\mu_i}{d t}=\mu_i (r_a-r_f)\big(1-\frac{A_{astro}}{\mu_i}\big) $ & $\frac{d \sigma_i}{d t} =2\sigma_i(r_a-r_f)-2\mu_i(r_a-r_f)A_{astro}+(r_a+r_f) \mu_i -2\mu_i(r_a-r_f)(\mu_i-A_{astro})$ \\
\\
~$sM3$  & $\frac{d\mu_i}{d t}=\mu_i (r_a-r_f)\big(1-\frac{A_{astro}+\gamma}{\mu_i+\gamma}\big)$   & ~ $\frac{d \sigma_i}{d t} =2\sigma_i(r_a-r_f)+\mu_i(r_a+r_f)-2\sigma_i(r_a-r_f)\big(\frac{A_{astro}+\gamma}{\mu_i+\gamma}\big)-2\mu_i(r_a-r_f) \big(1-\frac{A_{astro}+\gamma}{\mu_i+\gamma}\big)$  \\
\hline
\end{tabular}
\end{table}
To solve the differential equations (i.e., Eqs. \ref{s1}-\ref{ast3}) using Bayesian inference, we need to first specify the prior distributions for the unknown parameters (i.e., $r_a$, $r_f$, and $A_{astro}$), and then use Bayesian methods to update these priors based on the observed data. Let’s initially assume that the priors for all the parameters are independent and normal. This fairly non-informative prior allows for a wide range of possible values for the parameters. Next, to find the parameters of the developed stochastic models, the maximum likelihood estimation approach has been used \cite{frohlich2016inference}. The likelihood function implicitly assumes that the measured mean and variance of the data at each time point $t_n$ are normally distributed around the mean $(\mu_i(t_n,\theta))$ predicted by the model and variance $(\sigma_i(t_n,\theta))$ with the estimated mean and variance given by $\hat{\mu}_i(t_n,\theta)$ and $\hat{\sigma}_i(t_n,\theta)$. Moreover, the standard deviations for each distribution of mean and variance are functions of unknown parameter $\theta$ and are given by $Z_{\mu_i}(\theta)$ and $Z_{\sigma_i}(\theta)$. The methods used to develop this model were proposed by Fröhlich and colleagues respectively \cite{frohlich2016inference}. Therefore, the likelihood function is given as follows \cite{johnson2019cancer}:
\begin{equation}\label{l1}
    L(\theta)=\prod\limits_{i=1}^n \frac{1}{\sqrt{2\pi Z^2_{\mu_i}(\theta)}}exp\Big(-\frac{1}{2}\Big(\frac{\mu_i(t_n,\theta)-\hat\mu_i(t_n,\theta)}{Z_{\mu_i}(\theta)}\Big)^2\Big)\times \prod\limits_{i=1}^n\frac{1}{\sqrt{2\pi Z^2_{\sigma_i}(\theta)}}exp\Big(-\frac{1}{2}\Big(\frac{\sigma_i(t_n,\theta)-\hat{\sigma}_i(t_n,\theta)}{Z_{\sigma_i}(\theta)}\Big)^2\Big),
\end{equation}
 where $n$ is the number of observations and the negative log-likelihood function is given accordingly as \cite{johnson2019cancer}:
\begin{equation}\label{l2}
 NLL(\theta)=\frac{1}{2}\prod\limits_{i=1}^n\Big(\log 2\pi Z^2_{\mu_i}(\theta)+\Big(\frac{\mu_i(t_n,\theta)-\hat\mu_i(t_n,\theta)}{Z_{\mu_i}(\theta)}\Big)^2\Big)+\frac{1}{2}\prod\limits_{i=1}^n\Big(\log 2\pi Z^2_{\sigma_i}(\theta)+\Big(\frac{\sigma_i(t_n,\theta)-\hat{\sigma}_i(t_n,\theta)}{Z_{\sigma_i}(\theta)}\Big)^2\Big).
    \end{equation}
To measure overall time points, the above equations weigh equally to find the mean and variance of the data using the maximum likelihood parameter estimation method. We used the fminsearch function in MATLAB to find $NLL(\theta)$. In practice, Bayesian inference for the $A\beta$ model may require careful consideration of the choice of prior distributions and the appropriate likelihood function, as well as the use of appropriate computational methods for sampling from the posterior distributions. As a result, in this scenario, we may need to go to the next phase of the sensitivity analysis and examine the posterior density plots. Note that the posterior distributions for forecasting $A\beta$ dynamics change when alternative priors are specified. The posterior distribution of the intercept and residual variance of $A\beta$ will vary based on the supplied priors, indicating a
substantively evolving interpretation of the intercept depending on the priors. The systematic sensitivity analysis represents one of the areas for future studies.

\section{Analysis based on the experimental setup supported by ADNI data}\label{exp}
\subsection{ADNI data}
The datasets used in the preparation of this article were obtained from the Alzheimer's Disease Neuroimaging Initiative (ADNI) database \url{adni.loni.usc.edu}. The primary goal of the ADNI has been to test whether serial magnetic resonance imaging (MRI), positron emission tomography (PET), other biological markers, and clinical and neuropsychological assessment can be combined to measure the progression of AD. Specifically, we used the ADNI data prepared for the AD modelling challenge and incorporate the weak and strong effect of astrocytes into the deterministic and stochastic growth models of $A\beta$, and fitted the model for the progression of AD. Details on this are provided next.
\subsection{Modelling dynamics of participants} 
We use the quantitative template (available at \url{(https://adni.loni.usc.edu/)} as a CSV file) for the progression of the AD project data set, which includes the three ADNI phases: ADNI 1, ADNI GO, and ADNI 2. This dataset contains measurements from brain MRI, PET, Cerebrospinal Fluid Analysis (CSF), cognitive tests, demographics, and genetic information \cite{bilgel2019predicting,shaheen2023data}. From ADNI 1/GO/2, we used the data for $1706$ individuals with $6880$ visits. The details of patients' visits are provided in the template.  The ADNI Conversion Committee made clinical diagnoses of MCI (mild cognitive impairment), NL (normal), EMCI (early mild cognitive impairment), LMCI(late mild cognitive impairment), and AD based on the standards outlined in the ADNI protocol \cite{bilgel2019predicting,ghazi2021robust}. Following the ideas of \cite{bilgel2019predicting,shaheen2023data}, we designed the data for the individuals with clinical follow-up visits to fall within the baseline to $24$-month window. Moreover, clinical follow-up visits with improperly arranged dates are discarded for each individual. Also, based on the data statistics, an acceptable time gap of $12$ months is estimated for the present study. In the data set, measurements and clinical diagnoses with missing dates and information per visit are set to missing values in order to use the ABC method, and participants with fewer than two distinct visits are excluded. Notably, data sets containing missing values and clinical status are denoted as ''missing". For the present study, we have used only the AD patients' data (although there are also NL, EMCI, LMCI, and MCI data sets). We found that the participants with AD brains for each visit between $2$ years have different $A\beta$ concentrations as provided in the ADNI data set. $A\beta$ is a promising biomarker that is measured in CSF fluids collected from ADNI participants measured in $\si{mL}$. In our study, we investigated the concentrations of $A\beta$ in AD patients across a range of ages spanning from 50 to 90 years. Our analysis revealed that AD progression is more pronounced as age increases. Furthermore, we observed that a weak influence of astrocytes can actually promote the growth of $A\beta$. These findings provide valuable insights into the complex relationship between age, astrocytes, and the development of AD, which could aid in the development of effective treatment strategies. In our research, we used the mean values of $A\beta$ concentrations measured at $bl$, $12$ months, and $24$ months as the initial values of $A\beta$, which were set to $A\beta_{\rm bl} (57) =200mL, A\beta_{\rm {12-month-visits}} (57) =400mL ,A\beta_{\rm 24-month-visits} (57) =600mL$. It's important to note that the initial values of $A\beta$ were calculated as the average of the measurements taken at those three time points. However, in presenting the results, we reported the $A\beta$ concentration values in relation to the patient's age, which was measured in years. This allowed us to more accurately assess the relationship between $A\beta$ concentration and age over the course of the study period. Interestingly, as the patient's age increases, there is a rapid growth in $A\beta$ concentrations during the disease state which leads to the progression of AD. However, the influence of astrocytes behaves differently on the growth of $A\beta$. We are focusing on the evolution of $A\beta$ because there are many plaques composed of $A\beta$ in the AD patient's brain \cite{zhang2020mathematical,shaheen2023data}.

Importantly, inverse problems refer to the process of determining the input parameters or conditions that are responsible for producing a measured output \cite{xu2021solving}. To clarify, the stochastic growth models of $A\beta$ were used to generate synthetic data, which we then used as inputs to our inverse problems ($sM1-sM3$). In our case, we used Bayesian inference to determine the unknown input parameters that were responsible for generating the observed $A\beta$ concentration data. The measured output in our analysis was the $A\beta$ concentration data obtained from ADNI. The input we tried to cover were the parameters of the stochastic growth models of $A\beta$, which were used to generate synthetic $A\beta$ concentration data (i.e., $r_{a}$,$r_f$ and $A_{astro}$. For our analysis, using Bayesian inference, we feed the $A\beta$ concentrations data from ADNI to our models ($sM1-sM3$). We estimate the parameters accordingly using the maximum likelihood estimation as discussed in Section \ref{abc}. The three different proposed stochastic models explain significant changes in the dynamic behaviour of  $A\beta$. First, we look at whether $A\beta$ development in vitro is guided by fitting the simulated and ADNI data to the exponential growth model ($sM1$), which is often used to describe AD progression well below carrying capacity \cite{whittington2018spatiotemporal}. Then, we investigate the strong and weak astrocyte effects by fitting the simulated and ADNI data into the developed models (i.e., $sM2-sM3$). Details are given in the next Section.

\section{Results}\label{res}
The temporal probabilistic distribution of $A\beta$ provides valuable information for studying the progression of AD and developing effective treatments through the ADNI database. In this section, we will present the results obtained from the developed stochastic models presented in Section \ref{stoch}. In AD, the accumulation of $A\beta$ plaques is known to be a key pathological feature of the disease. The spatial distribution of these plaques throughout the brain is known to be heterogeneous, with some regions being more affected than others. Additionally, the temporal progression of the accumulation of $A\beta$ plaques can also vary, with some patients showing a more rapid progression of the disease than others. The ADNI database contains data on patients with varying degrees of AD, and as a result, the $A\beta$ accumulation patterns may differ across different patients. Moreover, the temporal aspect of $A\beta$ accumulation in AD refers to the progression of the disease over time, as the accumulation of  $A\beta$ and the formation of plaques gradually spread throughout the brain, and this progression can vary among individuals. Our models produce the temporal evolution of $A\beta$ using simulated data and the ADNI data.

\begin{figure}[htbp]
     \centering
     \begin{subfigure}[b]{0.48\textwidth}
         \includegraphics[width=\textwidth]{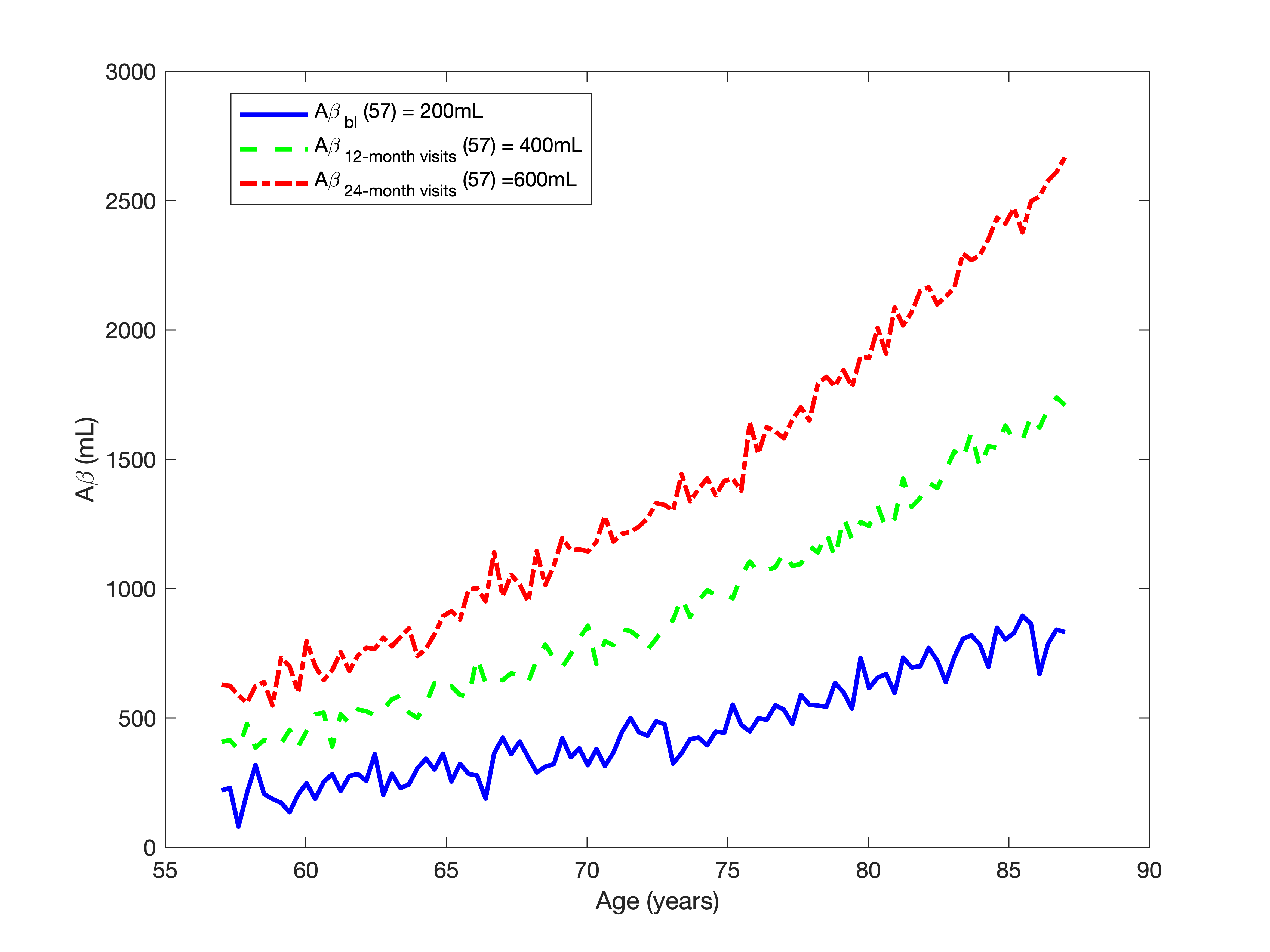}
         \caption{}
   \label{a}
     \end{subfigure}
     \begin{subfigure}[b]{0.48\textwidth}
         \includegraphics[width=\textwidth]{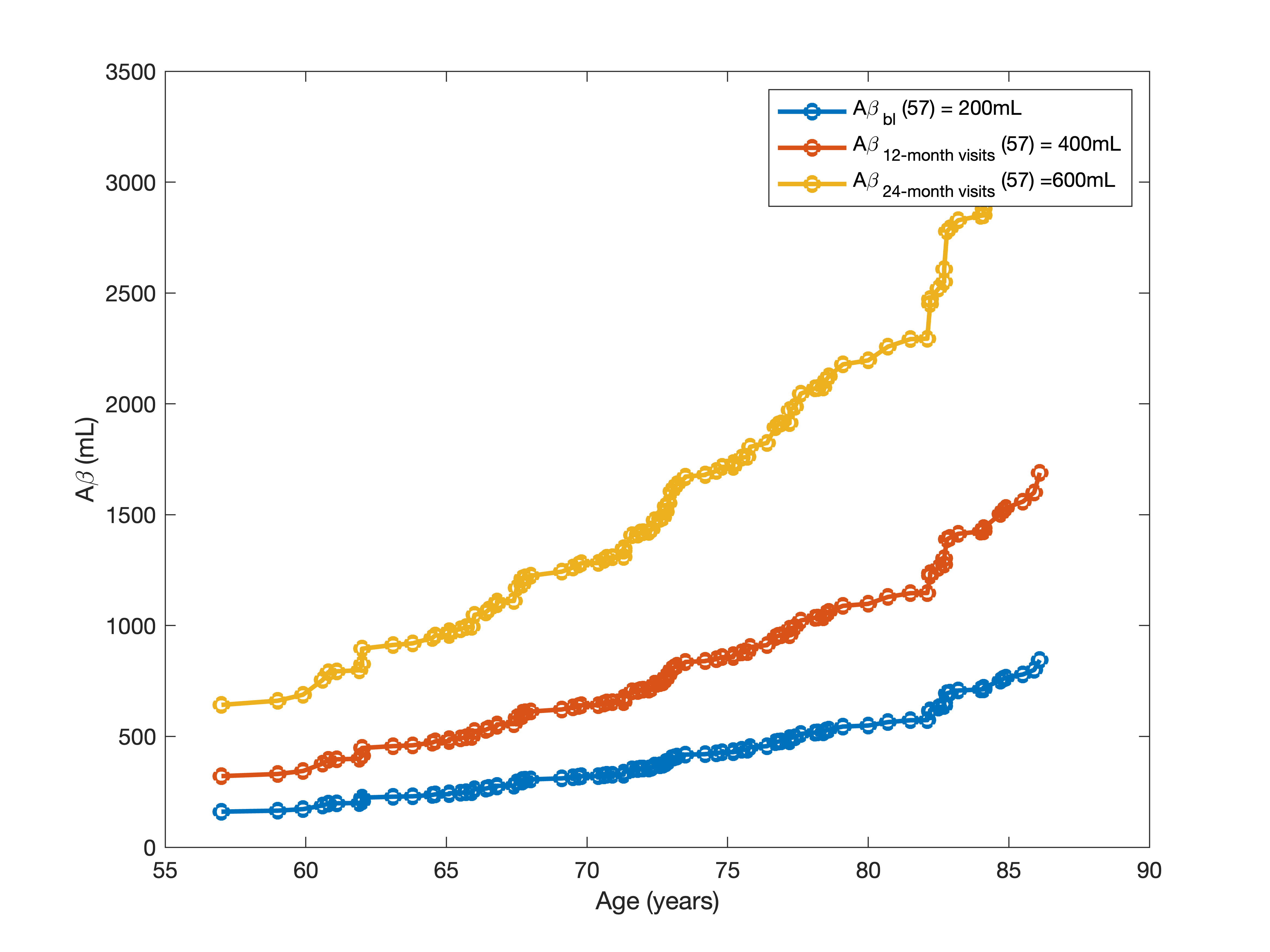}
         \caption{}
       \label{b}
     \end{subfigure}
    \caption{(Color online) (a) The trajectories of $A\beta$ growth model (i.e., $sM1$ given in Table \ref{tab:1}) with simulated data output from $10,000$ simulations with random noise, (b) The trajectories of $A\beta$ growth model with fitted ADNI data, the three curves show the fitted $A\beta$ data at $bl$ (blue), $12$ months (red) and $24$ months (yellow) with respect to age. $A\beta$ is a promising biomarker that is measured in ($\si{mL}$) in ADNI fluids such as in CSF \cite{bilgel2019predicting}.
    \label{fig:3}}
\end{figure}
The first goal is to plot the predicted results by fitting the simulated data before adding the ADNI data. Therefore, to fit and generate trajectories of the simulated data that resembled the ADNI data, we incorporated a random noise term. For instance, a random noise term was added to the growth model of $A\beta$ (i.e., $sM1$ presented in Table \ref{tab:1}) and the simulated trajectories of this model are presented in Fig. \ref{fig:3} (a) with initial values of $A\beta$ (i.e., $[200; 400; 600]$). The initial values of $A\beta$ were taken at the time points corresponding to the baseline, $12$-month, and $24$-month visits to align the simulated data with the ADNI data. The time vector span is defined as linspace$(0, 30, 100)$, which means that the ODE is solved for $100$ time points between $t=bl$ months and $t=24$ months (i.e., a 3-year period).  Moreover, random noise with a standard deviation of $50$ is added to the simulated trajectories to account for variability in the data as shown in Fig. \ref{fig:3} (a). The results are plotted as a function of age in years, assuming that the participants are all $57$ years old at baseline in Fig. \ref{fig:3} (a). The simulated data and fitted ADNI data can be connected by using the ABC technique to fit a model to the simulated data, and then generating predictions such as the temporal trajectories for ADNI data based on that model. Next, the ADNI data have been fitted to the growth model of $A\beta$ (i.e., $sM1$) presented in Table \ref{tab:1}). In addition, to get a temporal distribution of $A\beta$ growth and to plot the temporal trajectories at each initial condition, we fitted the ADNI data using the ABC method independently as shown in Fig. \ref{fig:3} (b). At first, the stochastic trajectories of $A\beta$ were sampled at $bl-24$ month for the stochastic model (i.e., $sM1$ presented in Table \ref{tab:1}), corresponding to the ages utilized in the experimental measurements of $A\beta$. The average $A\beta$ trajectories for the fitted ADNI data are depicted in Fig. \ref{fig:3} (b) with the initial conditions $A\beta_{\rm bl} (57) = 200\si{mL}$; $A\beta_{\rm 12-month-visits} (57)= 400\si{mL}$, $A\beta_{\rm 24-month-visits} (57)=600\si{mL}$ for AD patients. Note that the initial concentrations are calculated (based on the data given on the template) as the mean of $A\beta$ at the baseline visit (i.e., $A\beta_{\rm bl} (57) =200\si{mL}$), the mean of $A\beta$ at the $12$ month visit (i.e., $A\beta_{\rm 12-month-visits} (57) =400\si{mL}$), and the mean of $A\beta$ at the $24$ month-visit (i.e., $A\beta_{\rm 24-month-visits} (57) =600$) and the results are plotted for $A\beta$ against the age of patients. The time used for the simulation and plotting of results spans $30$ years, with time points spaced evenly over this interval.  Moreover, in Fig. \ref{fig:3} (b), the blue curve shows the fitted $A\beta$ data at baseline visits, the red curve shows the fitted $A\beta$ data at $12$ months visits and yellows shows the fitted $A\beta$ data at $24$ months visits with respect to the ages of the AD patients.

The patterns observed in the fitted ADNI data for $A\beta$ concentration are the same as observed in the growth of $A\beta$ concentration presented in Fig. \ref{fig:2} (a) without the influence of astrocytes. Hence our idea of developing the deterministic models before stochastic models were helpful. The disease progression of Alzheimer's is often associated with an increase in the concentration of $A\beta$ in the brain. This is particularly evident as patients age, with rapid growth in $A\beta$ concentrations observed over time. In Fig.\ref{fig:3} (b), we can clearly see this trend, where the $A\beta$ concentrations measured during the three visits are higher as patients get old. This suggests that the disease continues to progress, leading to an accumulation of $A\beta$ in the brain as the patient gets older. For instance, at the baseline visits ( blue curve), we see that in the patients whose age ranged from $50-90$, there is an increase in the concentration of $A\beta$ as they get older. Hence, as age increases, the disease progresses more in AD patients.

\begin{figure}[htbp]
    \centering
   \includegraphics[scale=0.5]{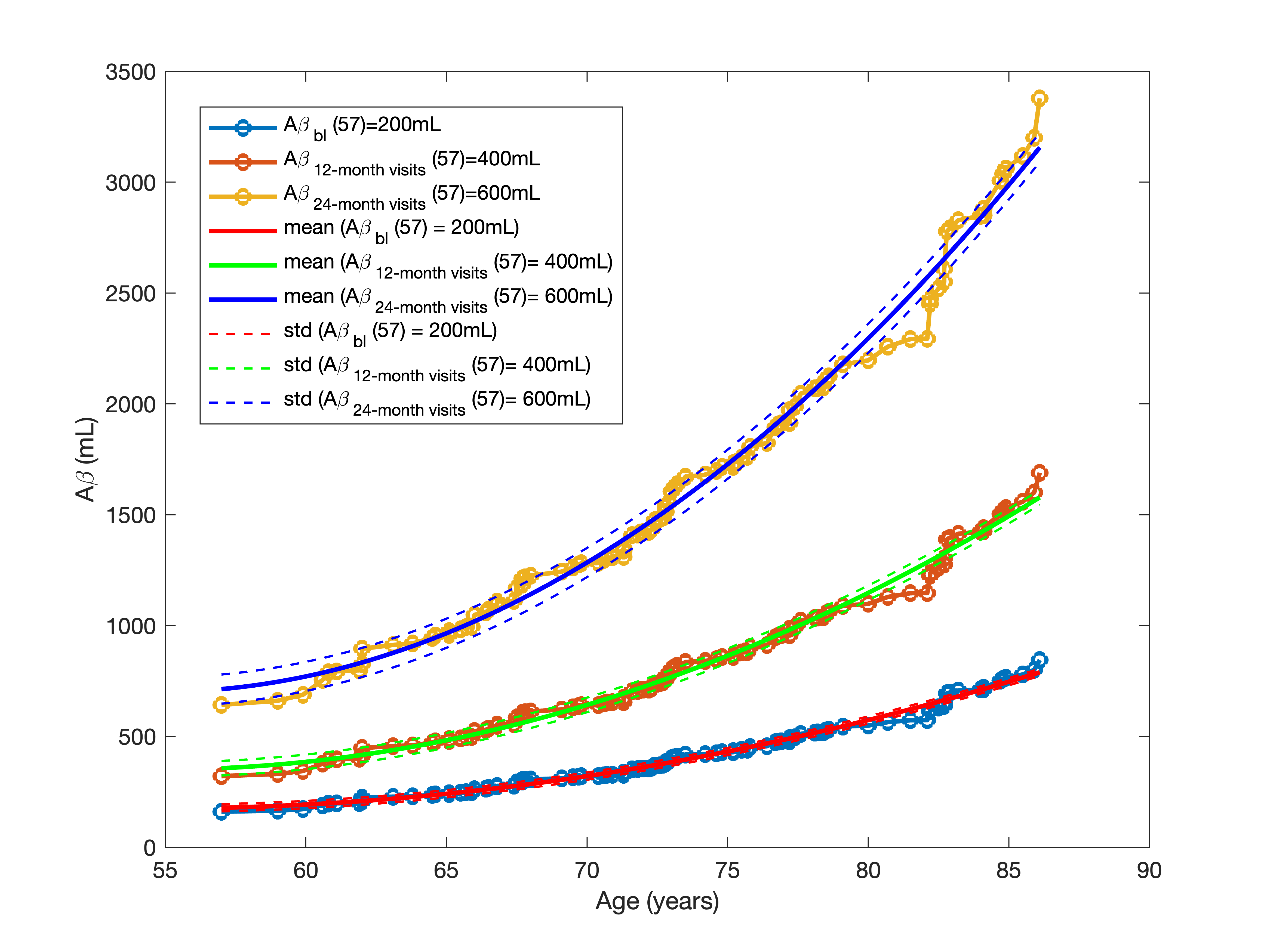}
    \caption{(Color online) The temporal distribution of $A\beta$ growth and the mean and variance of fitted ADNI data for AD patients. The three curves show the fitted $A\beta$ data from 3 visits at $bl$, $12$ months and $24$ months with respect to the age of the simple $A\beta$ growth model ($sM1$ presented in Table \ref{tab:1} and the corresponding mean and variance presented in Table \ref{tab:2}) using the ABC technique.}
    \label{fig:4}
\end{figure}
 We plot the results in Fig. \ref{fig:4}, using a standard deviation of $\pm 2$ mean, knowing that in any distribution, about $95\%$ of values will be within $2$ standard deviations of the mean. Still, we observed a notable shift in the measured data around the age of $82$ years. This shift may indicate a potential bias or uneven distribution specifically centred around that age, resulting in a distinct kink in the plotted data. It is possible that there are fewer data points or a specific subset of data points available around that age, leading to a noticeable deviation. However, despite this anomaly, the model demonstrates its capability to explain a substantial portion of the observed variability in the data. To plot the mean and variance of fitted data, we first need to fit a curve to the data. Fig. \ref{fig:4} shows the curve for the mean and variance of the fitted ADNI data for the simple $A\beta$ growth model ($sM1$). It can be seen clearly that the data fit well according to the mean and variance.  In Fig. \ref{fig:4}, we can see that the mean of the fitted data follows the trend of the ADNI data reasonably well, with both showing a gradual increase in $A\beta$ concentration with age. Additionally, we can see that the variance of the fitted data is also consistent with the ADNI data, with both showing an increasing trend with age. This provides a good fit which means that the fitted curve is able to capture the general trend in the data, as well as the variability around the trend. In other words, the model is able to explain a significant portion of the observed variability in the data. This is an important finding as it suggests that the $sM1$ model provides a good representation of the underlying biological process that governs the accumulation of $A\beta$ in the brain over time. The results suggest that the occurrence of $A\beta$ deposits in AD patients is a dynamic process with significant growth. When $A\beta$ deposits are created in the brain, they can become more stable and resistant to degradation. This is due to fibrillogenesis, where individual $A\beta$ peptides aggregate and form stable fibrils \cite{thompson2020protein}. The formation of these fibrils is thought to be a key step in the development of $A\beta$ plaques, which are the hallmark of AD. Our results are consistent with other studies demonstrating a significant increase in $A\beta$ deposition throughout the preclinical stages of AD \cite{sojkova2011longitudinal,villemagne2011longitudinal,vlassenko2011amyloid,shaheen2023data}. This information on the growth of $A\beta$ concentration may be useful in designing potential AD treatment strategies \cite{mosconi2018increased,insel2020neuroanatomical}.

The experimental growth measurements at low initial $A\beta$ concentrations reflect the variety in observed $A\beta$ trajectories for a single initial condition \cite{bertens2015temporal}. Importantly, given the inherent stochasticity of the fragmentation and aggregation processes, which is evident at low initial values of $A\beta$ examined in this study, our variation in $A\beta$ growth dynamics is to be anticipated (Fig. \ref{fig:4}). Because stochasticity is more visible and may be detected experimentally at low $A\beta$ concentrations (Fig. \ref{fig:4}), such dynamics are best represented by a stochastic rather than a deterministic process. Therefore, the study concludes that the simple $A\beta$ growth model ($sM1$) is a suitable model for predicting $A\beta$ trajectories in the context of AD. Further, to see if the preliminary observations of growth-rate scaling with initial $A\beta$ could be explained by alternative growth models that account for the astrocyte effect ($sM2-sM3$), the experimental data of ADNI growth trajectories for $A\beta$ were calibrated to the stochastic models using the stochastic modelling framework shown in Fig. \ref{fig:1} as follows.

\begin{figure}[htbp]
     \centering
     \begin{subfigure}[b]{0.48\textwidth}
         \includegraphics[width=\textwidth]{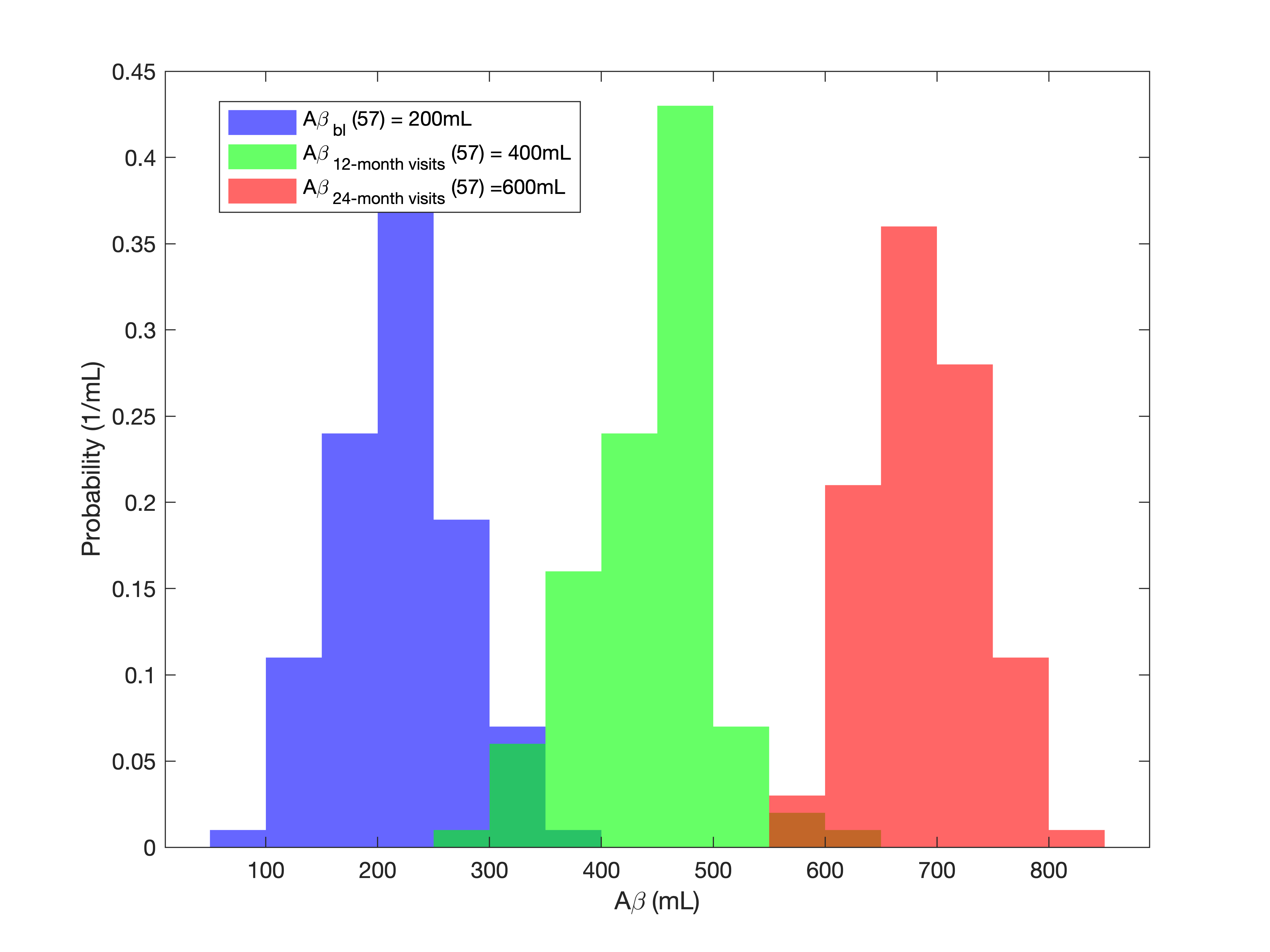}
         \caption{}
   \label{a}
     \end{subfigure}
     \begin{subfigure}[b]{0.48\textwidth}
         \includegraphics[width=\textwidth]{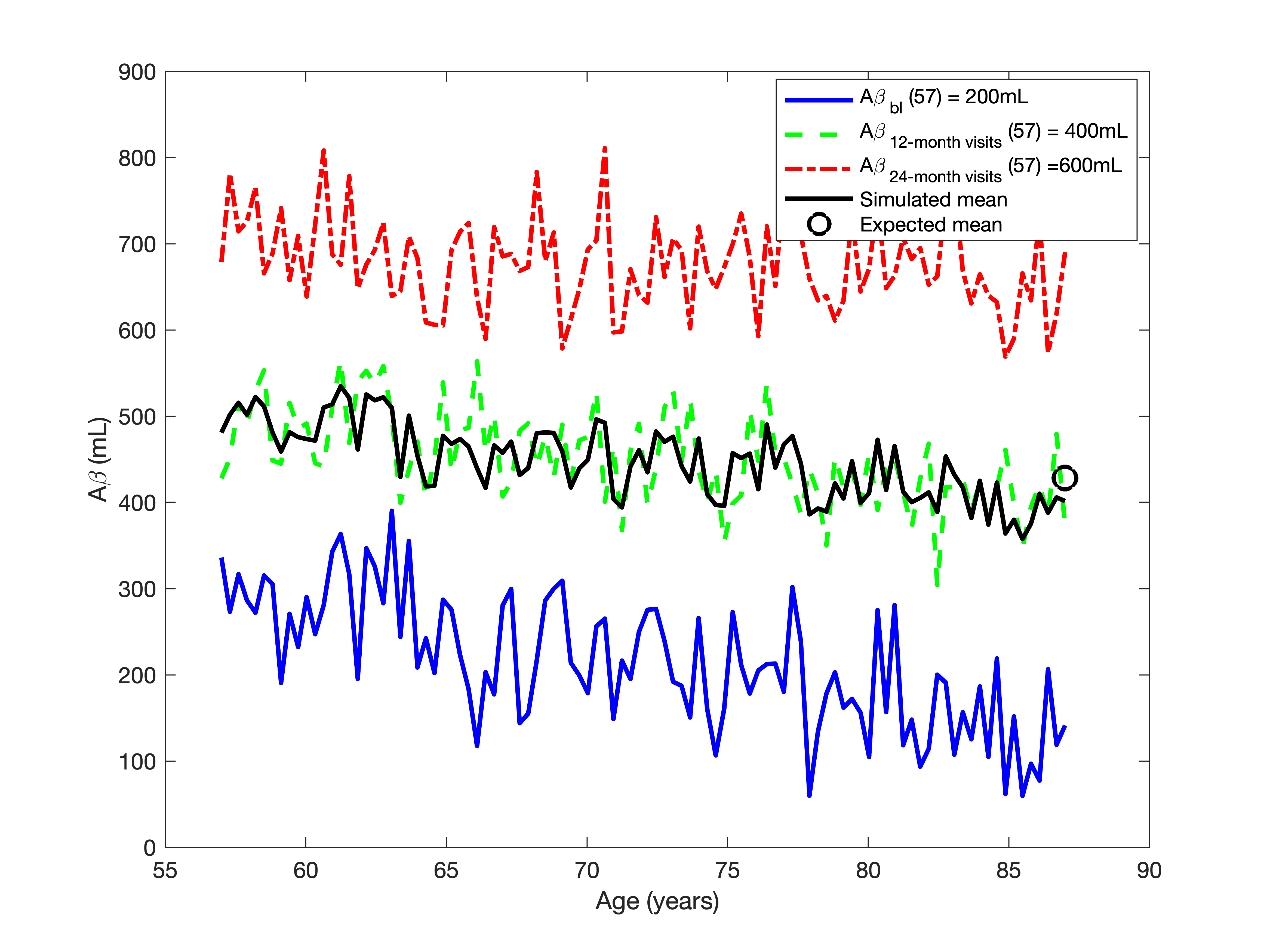}
         \caption{}
       \label{b}
     \end{subfigure}
        \caption{(Color online) The (a) histograms with respect to trajectories of simulated stochastic $A\beta$ concentrations ($\si{mL}$) and (b) the trajectories with simulated data of strong astrocyte effect model ($sM2$ presented in Table \ref{tab:1}), output from $10,000$ simulations with $A\beta_{\rm bl} (57)=200\si{mL}$, $A\beta_{\rm 12-month-visits} (57)=400\si{mL}$, $A\beta_{\rm 24-month-visits} (57)=600\si{mL}$, $r_a=0.002413,r_f=0.0006536, A_{astro}=700\si{mL}$ from 3 visits at $bl$ (blue), $12$-months (green) and $24$-months (red), with respect to age. The simulated mean (black line) and expected mean (black circle) present the best-fit mean at $A\beta_{\rm 12-month-visits} (57)=400\si{mL}$.}
        \label{fig:5}
\end{figure}
Next, before fitting the ADNI data to the strong and weak astrocyte effect model, we fitted the simulated data first. The results shown in Fig. \ref{fig:5} (a-b), present the effects of adding random white noise to the strong astrocyte effect model ($sM2$ presented in Table \ref{tab:1}). In Fig. \ref{fig:5} (a), the histograms are plotted using Eq. \ref{PDF} which shows the distribution of simulated stochastic $A\beta$ concentrations corresponding to patients with ages $57-87$ \cite{bertens2015temporal,bilgel2019predicting,ghazi2021robust}. The $x$-axis represents the range of values of the $A\beta$ concentrations being measured, and it is divided into a set of intervals or bins with respect to the initial conditions from 3 visits. The $y$-axis represents the frequency of observations falling within each bin. In the context of the simulation of stochastic $A\beta$ concentrations, the histogram shows the distribution of the concentrations that were simulated over $10,000$ trials, with starting concentrations of $A\beta$ i.e., $200\si{mL},400\si{mL},600 \si{mL}$. We can see that $ A\beta$ concentrations are normally distributed as depicted in Fig. \ref{fig:5} (a), therefore the simulated data fit well to the chosen model. Moreover, the simulated stochastic $A\beta$ trajectories for $A\beta$ from 3 visits are presented in Fig. \ref{fig:5} (b) demonstrating the predictable decline in growth dynamics of $A\beta$ that is visible at low initial levels of $A\beta$ since $A_{astro}=700\si{mL}$ is greater than the initial values of $A\beta$. Moreover, we sample time evenly from the stochastic simulated $A\beta$ trajectories and determine the simulated mean and expected mean of $A\beta$ at each time point for $A\beta_{\rm 12-month-visits} (57)=400\si{mL}$ at $12$ month visit as seen in Fig. \ref{fig:5} (b) with respect to age. The simulated data fit well as per the expected mean and simulated mean, according to the strong astrocyte effect model ($sM1$) on birth and death using the ABC technique of statistical approaches. As seen from Fig. \ref{fig:5} (b), the chosen model correctly predicted the simulated mean and expected mean of the simulated data. For instance, at $A\beta_{\rm 12-month-visits} (57)=400\si{mL}$ at $12$ month visit as seen in Fig. \ref{fig:5} (b) (black curve) the simulated mean and expected mean are similar or have little difference, thus it suggests that the model fits well with the simulated data. For the strong astrocyte effect model ($sM2$) given in Table \ref{tab:1} with mean given in Table \ref{tab:2}, if the concentration of astrocytes ($A_{astro}$) is greater than the initial concentration of $A\beta$, there is a decrease in the concentration of $A\beta$ as shown in Fig. \ref{fig:5} (b). This suggests that the strong astrocyte effect facilitates the clearance of $A\beta$. Note that the trajectories of the simulated stochastic $A\beta$ concentrations do not start exactly from the deterministic initial concentrations of $A\beta$ due to the inherent nature of stochastic simulations and the random processes involved.

\begin{figure}[htbp]
     \centering
     \begin{subfigure}[b]{0.48\textwidth}
         \includegraphics[width=\textwidth]{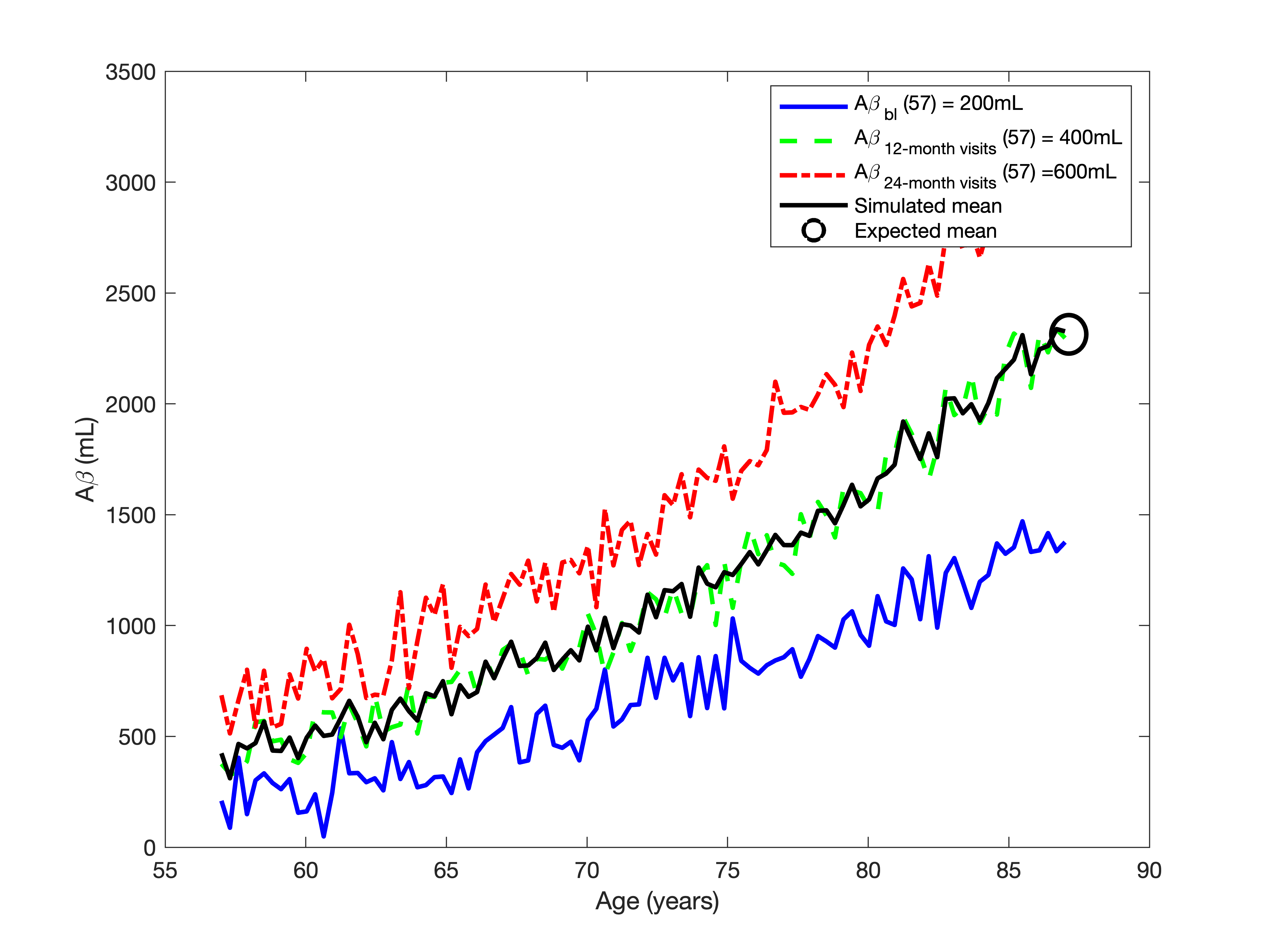}
         \caption{}
   \label{a}
     \end{subfigure}
     \begin{subfigure}[b]{0.48\textwidth}
         \includegraphics[width=\textwidth]{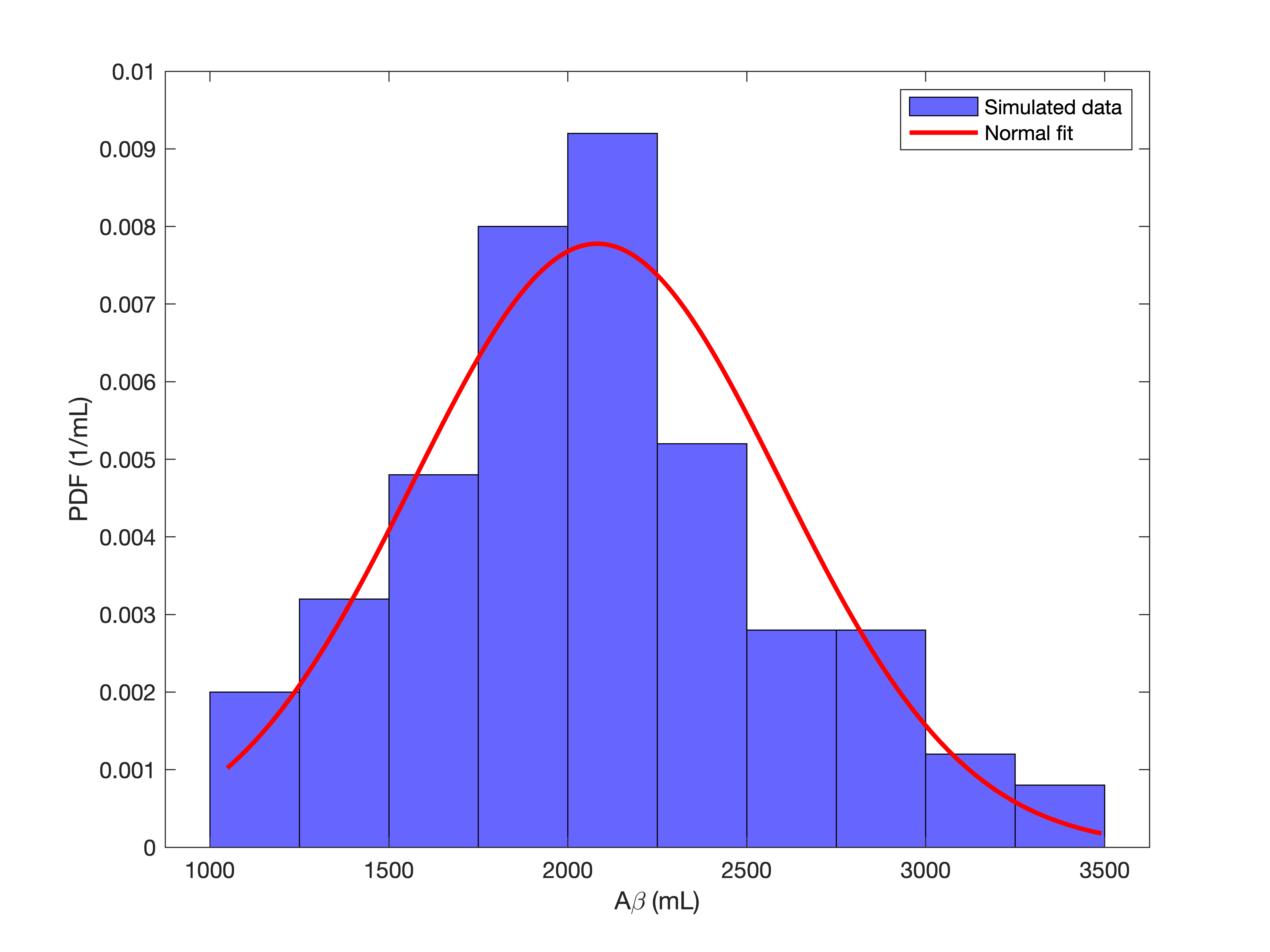}
         \caption{}
       \label{b}
     \end{subfigure}
 \caption{(Color online) (a) The trajectories of weak astrocyte effect model ($sM3$ presented in Table \ref{tab:1}) with simulated data, output from $10,000$ simulations with $r_a=0.002413,r_f=0.0006536,A_{astro}=100\si{mL}$ from 3 visits at $bl$ (blue), $12$ months (green) and $24$ months (red). The simulated mean (black line) and expected mean (black circle) present the best-fit mean at $A\beta_{\rm 12-month-visits} (57)=400\si{mL}$ and (b) the histograms of simulated stochastic $A\beta$ concentrations data ($\si{mL}$) and normal fit (red line) are presented at $24$ month visits with $A\beta_{\rm 24-month-visits} (57)=600\si{mL}$.}
   \label{fig:6}
\end{figure}
Similarly, Fig. \ref{fig:6} (a-b) shows the results of adding random white noise to the weak astrocyte effect model ($sM3$ presented in Table \ref{tab:1}) of $A\beta$. The simulated $A\beta$ trajectories from three visits with respect to age are shown in Fig. \ref{fig:6} (a). The model successfully fits the data well under initial conditions and throughout the whole time period, as shown by the best fit of the simulated mean and expected mean. The histogram plotted using Eq. \ref{PDF} shows that the simulated data fit well in this case also as shown in Fig. \ref{fig:6} (b) as per the normal fit. As can be seen in Fig. \ref{fig:6} (a), the weak astrocyte effect (where $A_{astro}=100\si{mL}$ is lower than the initial concentrations of $A\beta$) would result in an increase in the concentrations of $A\beta$. The trends of the weak and strong astrocyte effect models' fitting to the simulated data over time for the average growth of $A\beta$ are shown in Figs (\ref{fig:5}-\ref{fig:6}). The pattern indicates a rise in the net growth rate of $A\beta$ for the weak astrocytes effect model ($sM3$ presented in Table \ref{tab:1}). However, in Fig. \ref{fig:5} (a-b) for the strong astrocytes effect model ($sM2$ presented in Table \ref{tab:1}) the average growth of $A\beta$ is decreasing. This shows that the strong astrocyte effect has the greatest influence on the clearance $A\beta$. They help to slow down the growth of $A\beta$ which leads to reduced AD progression. Moreover, it was suggested that the weak astrocyte effect is an important regulator of the brain's inflammatory response and is a commonly recognized hallmark of AD \cite{frost2017role}. 

\begin{figure}[htbp]
     \centering
     \begin{subfigure}[b]{0.48\textwidth}
         \includegraphics[width=\textwidth]{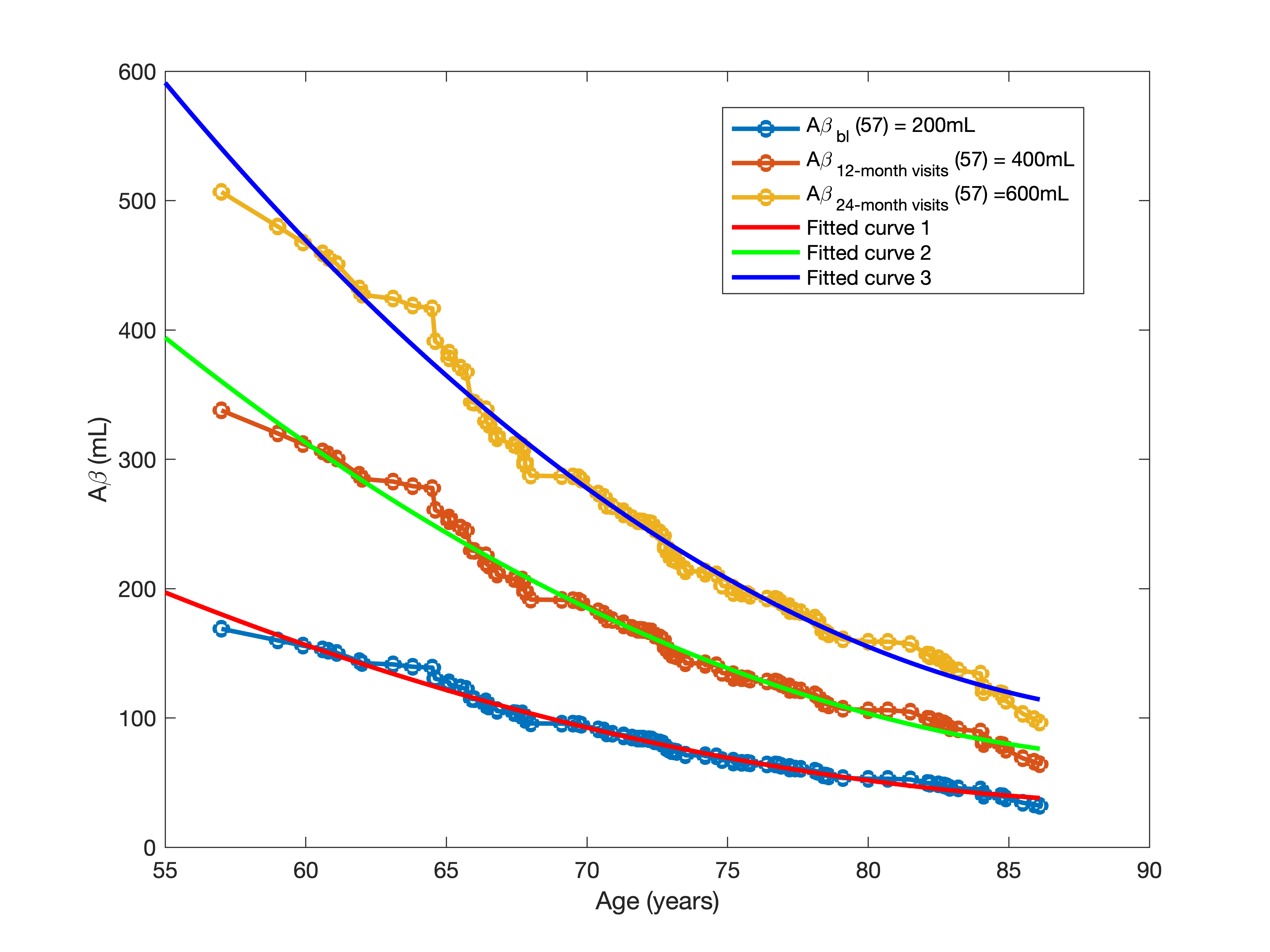}
         \caption{}
   \label{a}
     \end{subfigure}
     \begin{subfigure}[b]{0.48\textwidth}
         \includegraphics[width=\textwidth]{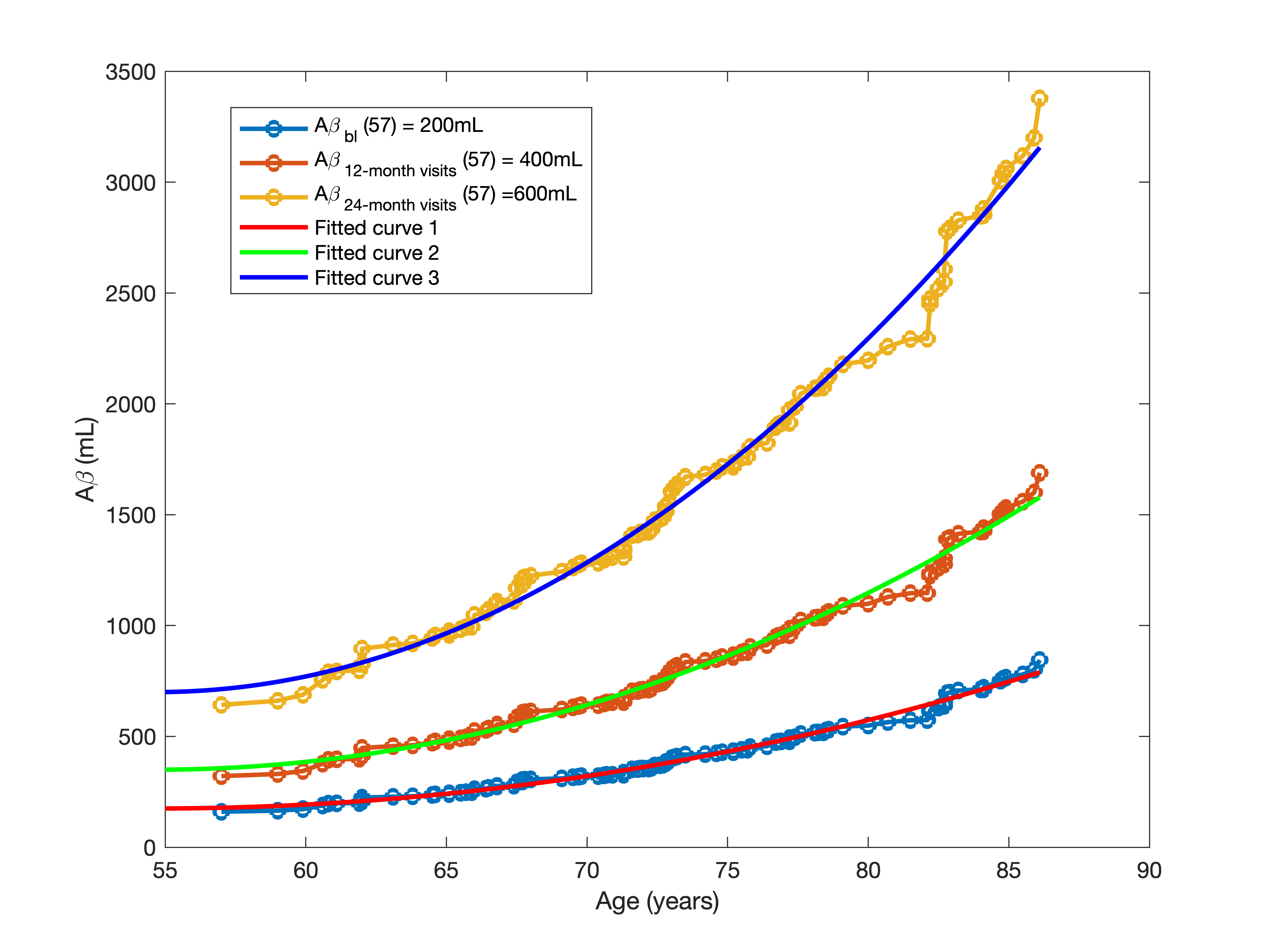}
         \caption{}
       \label{b}
     \end{subfigure}
    \caption{(Color online) The temporal distribution of $A\beta$ growth for strong and weak astrocytes effect models (i.e., $sM2-sM3$) and the mean and variance of fitted ADNI data for AD patients from $bl$ to $2$-years of visits as per patients age. The three curves show the fitted $A\beta$ data from 3 visits at $bl$, $12$ months and $24$ months for $A\beta_{\rm bl} (57),A\beta_{\rm 12-month-visits} (57),A\beta_{\rm 24-month visits} (57) =200 \si{mL},400\si{mL},600\si{mL}$ with respect to the age of the (a) strong astrocyte effect on $A\beta$ growth model ($sM2$),  $A_{astro}=700\si{mL}$ and (b) weak astrocyte effect on $A\beta$ growth model ($sM3$) presented in Table \ref{tab:1}, $A_{astro}=100\si{mL}$ and the corresponding fitted curves.}
        \label{fig:7}
\end{figure}
In Figs. (\ref{fig:5}-\ref{fig:6}) we fit the simulated data to the stochastic models ($sM2-sM3$) presented in Table \ref{tab:1}. Using the same variable values, the next goal is to fit ADNI data to these stochastics to reveal the astrocyte effect. Therefore, accordingly, we have fitted the ADNI data to the weak and strong astrocyte effect models as presented in Fig.\ref{fig:7} (a-b). We have fitted the curve for AD patients in between $bl$ to $2$-years of visits, and their ages ranged from $50-90$ years. Both models provided a good fit to the ADNI data of $A\beta$ as per the fitted curves. We can compare the behaviour of both models ($sM2-sM3$) with the best fit. The initial conditions for $A\beta$ are chosen here are $A\beta_{\rm bl} (57),A\beta_{\rm 12-month-visits} (57),A\beta_{\rm 24-month-visits} (57)=200\si{mL},400\si{mL},600\si{mL}$ are the means of $A\beta$ at baseline, $12$ months, and $24$ months visits with respect to the ages of AD patients. We see in the growth curve for $A\beta_{\rm bl} (57) ==200$ for the strong astrocyte effect model ($sM2$ given in Table \ref{tab:1}) as shown in Fig. \ref{fig:7} (a), is $A\beta$ concentrations is decreasing for the patients aged $50$-$70$ years until the $A\beta$ concentrations are less than $A_{astro}$, but we observed the opposite trend in Fig. \ref{fig:7} (b). Since, there is a tremendous increase in the concentration of $A\beta$ for the weak astrocyte effect model, as shown in Fig. \ref{fig:7} (b) as the patient's age increases. Moreover, it is depicted in Fig. \ref{fig:7} (a) that, in the presence of a strong astrocyte effect when initial $A\beta$ is less than $A_{astro}$, there is a rapid decrease in the $A\beta$ concentration, which means strong astrocytes effect helps to clear $A\beta$. Whereas, as known, the weak astrocytes' effect on $A\beta$ growth will encourage chronic inflammatory activities in the brain, which will lead to neuronal death or brain injury \cite{batarseh2016amyloid}. However, the strong astrocytes effect helps to reduce $A\beta$ growth. These findings imply and further confirm that $A\beta$-astrocyte interaction is pathologically harmful and promotes AD \cite{hao2016mathematical,frost2017role}. 

\begin{figure}[htbp]
     \centering
     \begin{subfigure}[b]{0.48\textwidth}
         \includegraphics[width=\textwidth]{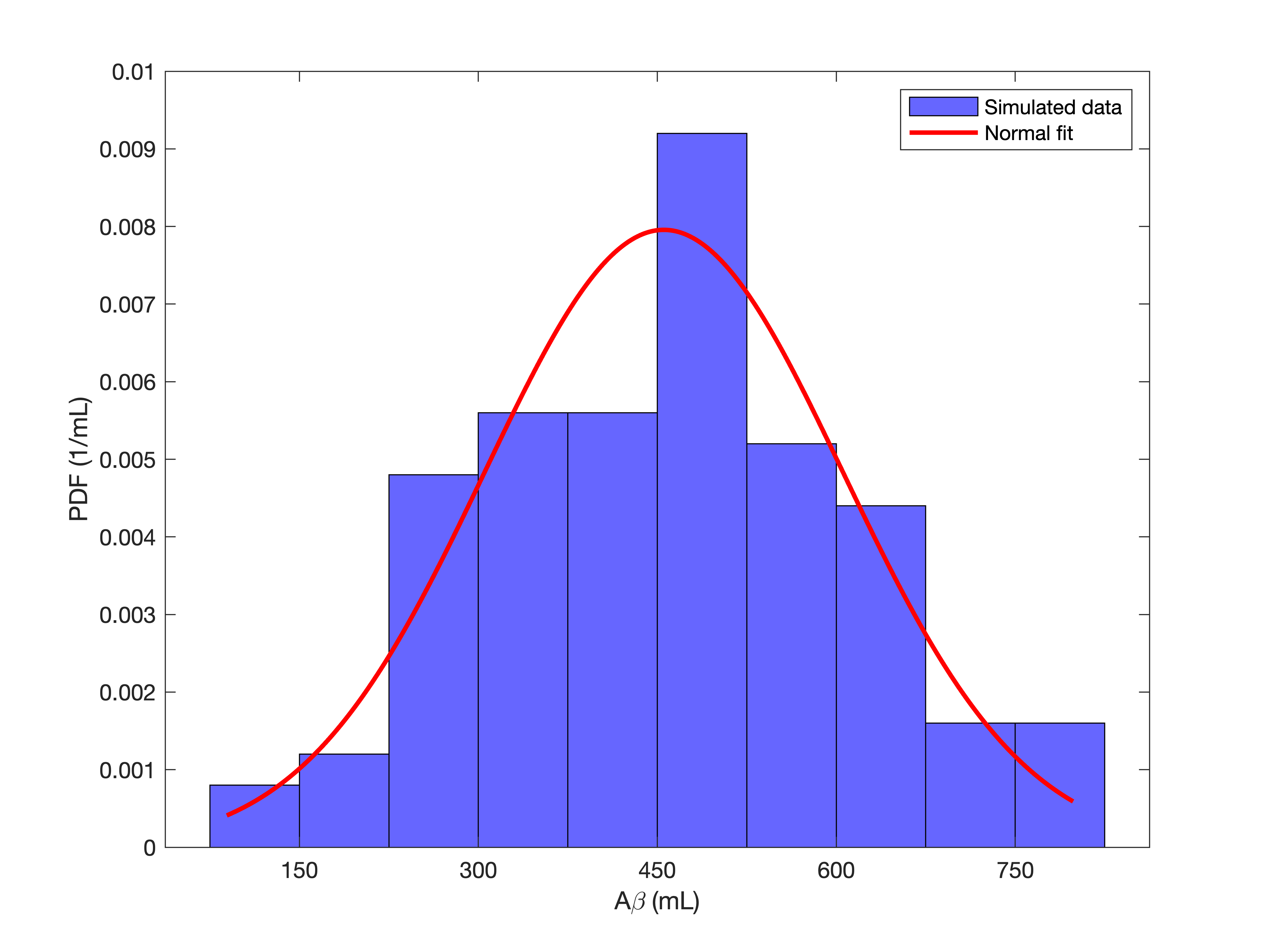}
         \caption{}
   \label{a}
     \end{subfigure}
     \begin{subfigure}[b]{0.48\textwidth}
         \includegraphics[width=\textwidth]{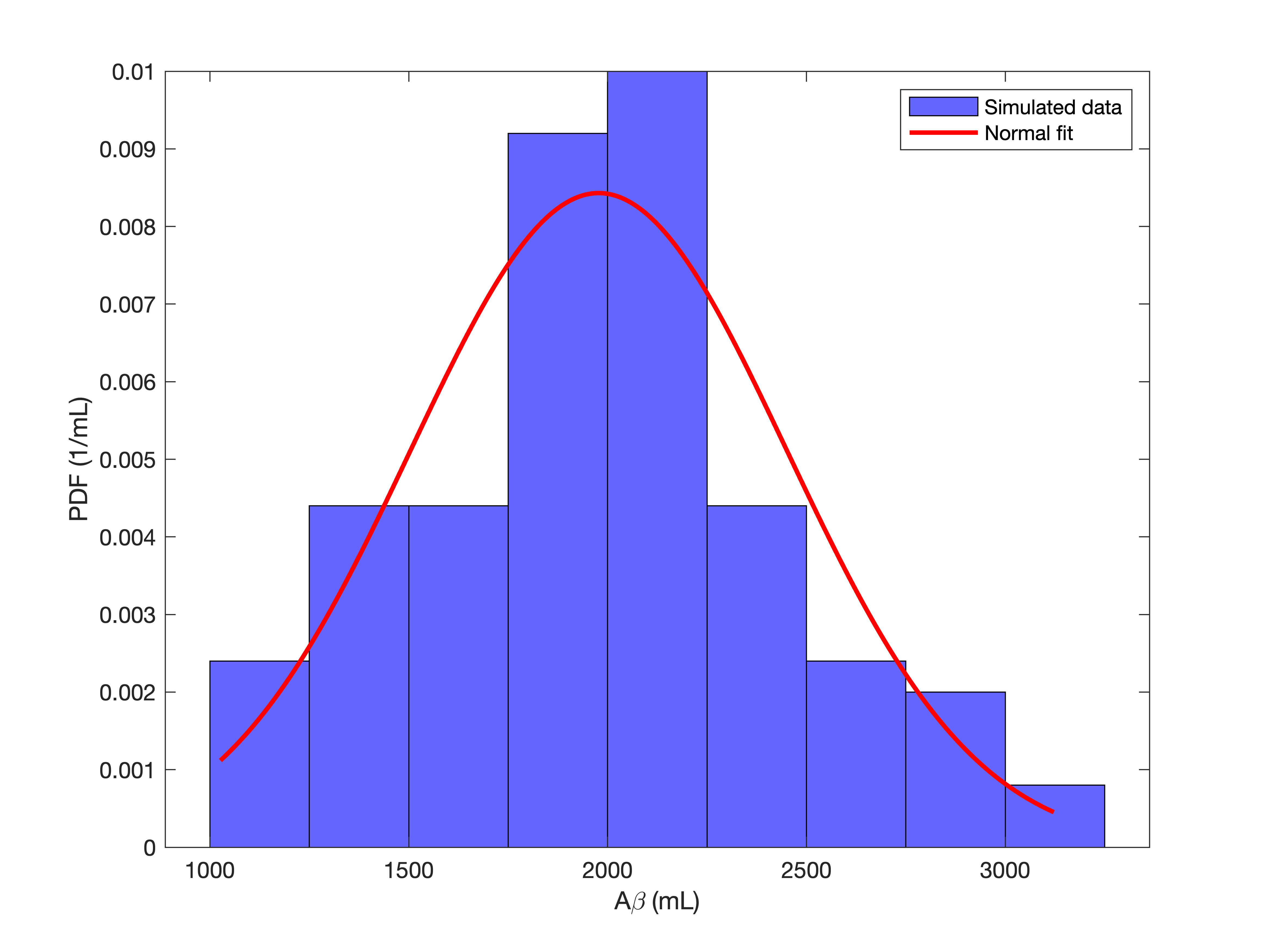}
         \caption{}
       \label{b}
     \end{subfigure}
        \caption{(Color online) Histograms of the fitted ADNI data for $A\beta$ to the (a) strong and (b) weak astrocyte effect models ($sM2$ and $sM3$ presented in Table \ref{tab:1}) using ABC technique for $A\beta_{\rm 24-month-visits} (57)=600\si{mL}$. The red lines represent the estimated normal distribution that best fits the data.}
        \label{fig:8}
\end{figure}
Moreover, the probability distribution function (pdf) has been estimated using Eq. \ref{PDF}, for both models, as shown in Fig. \ref{fig:8}(a-b) for $A\beta_{\rm 24-month-visits} (57)=600\si{mL}$. Also, we have fitted the ADNI data set (quantitative template) for AD patients as presented in Fig. \ref{fig:8} for strong and weak astrocyte models ($sM2-sM3$) given in Table \ref{tab:1}. The normal fit (red line) shows that the ADNI data fit well for $A\beta$ concentrations for both models presented in Fig. \ref{fig:8}. The motivation behind analyzing only the AD patients as opposed to NL, EMCI, etc, is that we can predict the weak and strong astrocyte effects on $A\beta$ growth in the diseased brain. Further, the results presented in Figs. (\ref{fig:7}-\ref{fig:8}) for the ADNI data set and the predicted $A\beta$ growth are in agreement with those estimated for the simulated data in Figs. \ref{fig:5}-\ref{fig:6} \cite{batarseh2016amyloid}. As we pointed out earlier, the stochastic nature of simulations and underlying random fluctuations result in slightly different starting concentration values that might be different from the deterministic case.

\begin{figure}[htbp]
     \centering
     \begin{subfigure}[b]{0.48\textwidth}
         \includegraphics[width=\textwidth]{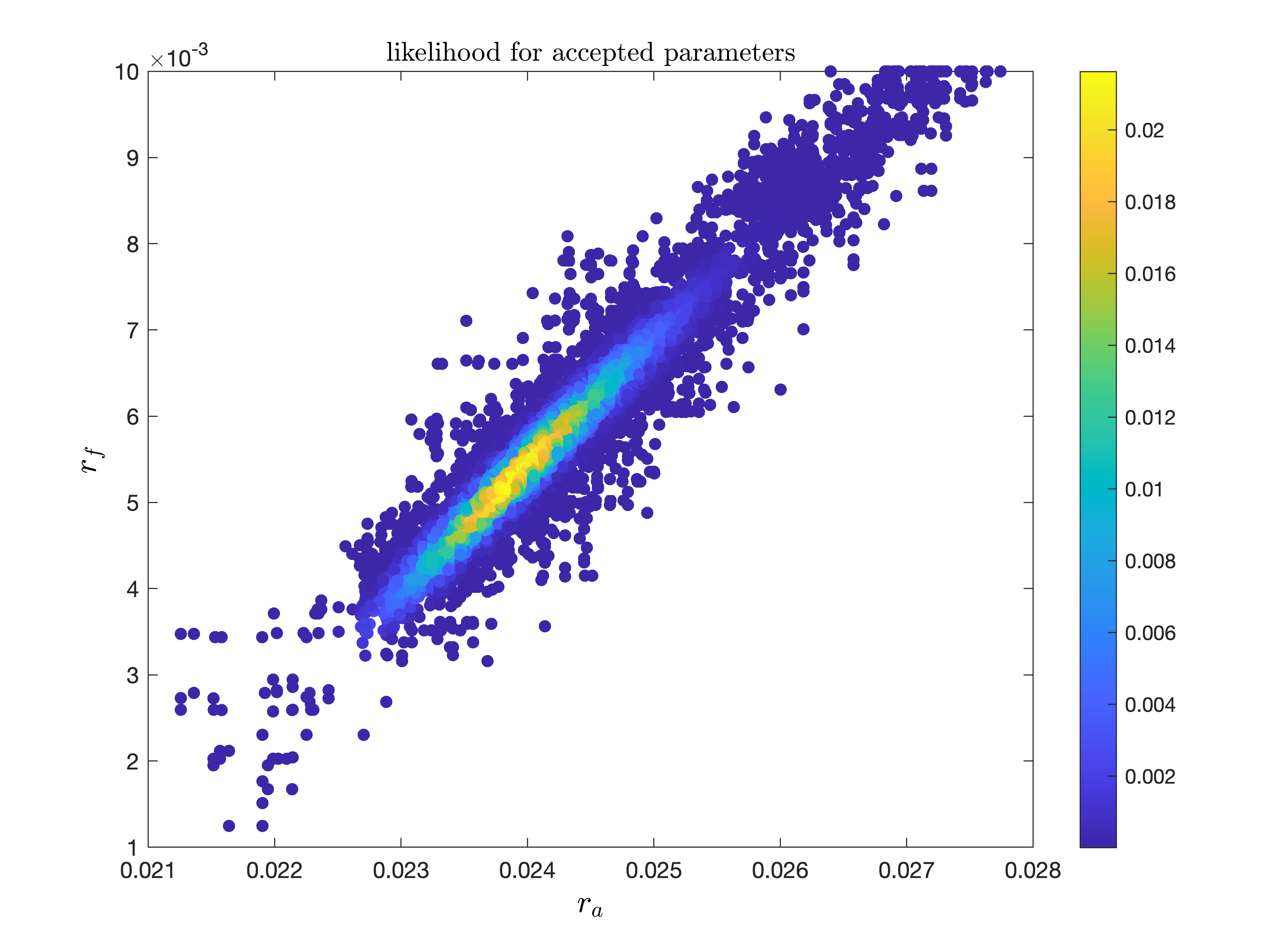}
         \caption{}
   \label{a}
     \end{subfigure}
     \begin{subfigure}[b]{0.48\textwidth}
         \includegraphics[width=\textwidth]{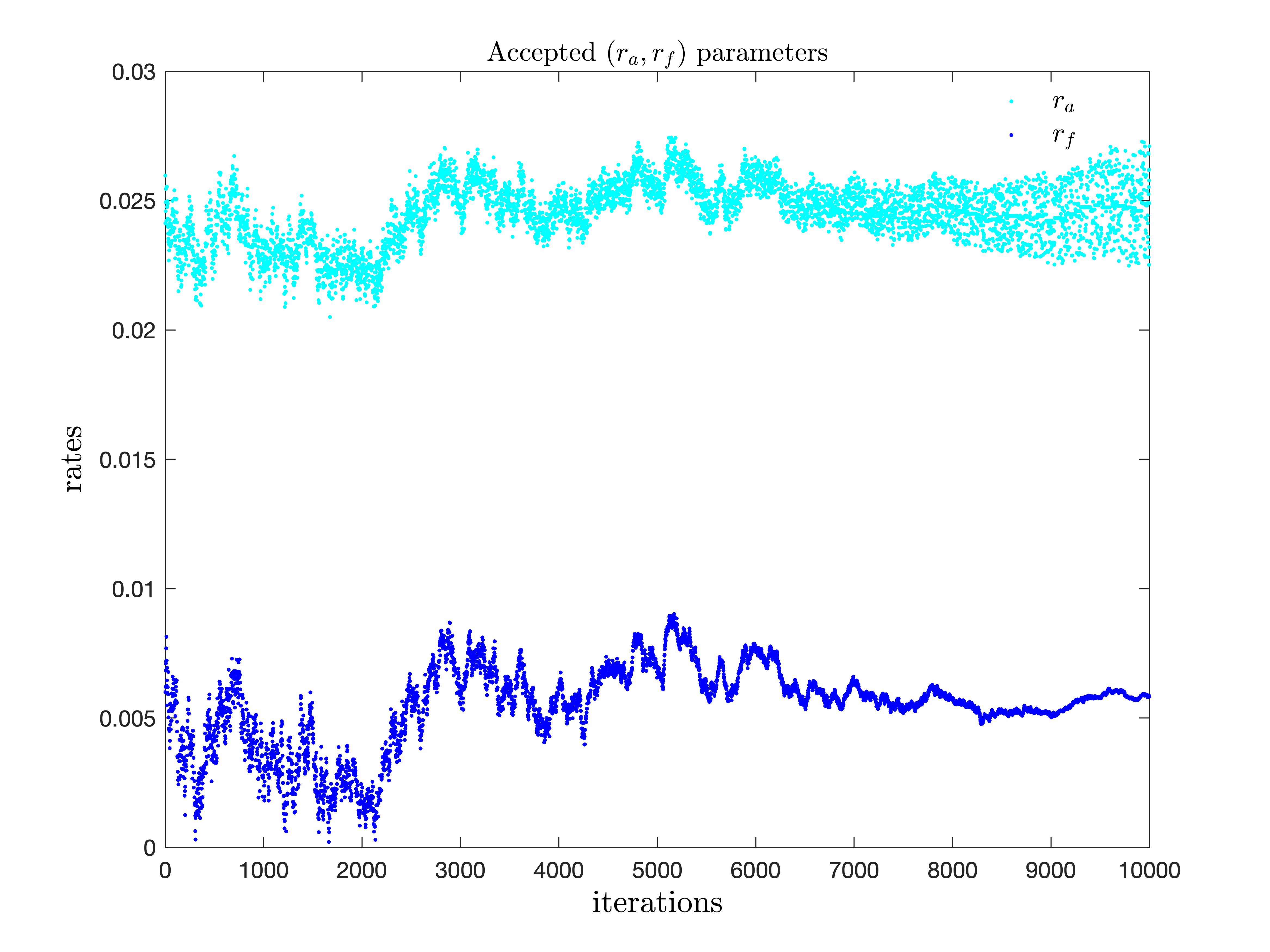}
         \caption{}
       \label{b}
     \end{subfigure}
        \caption{(Color online) (a)The parameters range examined is shown, and the parameter sets $r_a,r_f$ are colored according to likelihood estimation, showing that the ABC framework has converged on the true parameters, and (b) the rates represent the values of accepted parameters ($r_a,r_f$ ) for the ADNI data for a strong
astrocyte effect model ($sM2$).}
        \label{fig:9}
\end{figure}

Importantly, when it comes to fitting ADNI or simulated data, parameter estimation plays an important role. In the present study, we provide a stochastic modelling framework and ABC technique in Sections \ref{stoch}-\ref{abc} to distinguish predicted deviations for three stochastic models ($sM1-sM3$), and we calibrate the observed growth trajectories using the moment approach for stochastic parameter estimation. We perform parameter estimation using the technique of moments approach for each stochastic model (i.e., $sM1-sM3$ presented in Table \ref{tab:1}) \cite{frohlich2016inference}. This is done for the ADNI data set to test the hypothesis that our ABC framework offers an alternate $A\beta$ growth model with an astrocyte effect. Moreover, ABC and a  maximum likelihood parameter estimation technique (i.e., Eqs. \ref{l1}-\ref{l2}) were used to infer the parameters of the stochastic models \cite{beaumont2002approximate,calvetti2018inverse}. The goal of this technique is to find the values of the parameters that maximize the likelihood of the observed data. In other words, it is the method of finding the best set of parameters that can explain the data with the highest probability. In the present study for stochastic models (i.e., $sM1-sM3$ presented in Table \ref{tab:1}, the unknown parameters are $r_a,r_f$, and $A_{astro}$, and we fixed $\gamma$. We define a range of values for $r_a$ and $r_f$ that cover the plausible values of the parameters. For each combination of $r_a$ and $r_f$ values in the defined ranges, we calculated the log-likelihood of the observed data using the maximum likelihood estimation method. The likelihood function plot across $r_a$ and $r_f$ values, as depicted in Fig. \ref{fig:9} (a), can provide useful information about the quality of the parameter estimates and the fit of the model to the ADNI data.

Specifically, the likelihood function plot can show the range of parameter values that provide a good fit to the data and the sensitivity of the likelihood to changes in the parameter values. According to Fig. \ref{fig:9} (a), the parameter estimations for the strong astrocyte effect model ($sM2$) were found to be quite close to the true parameters (i.e., $r_a = 0.002413$, $r_f = 0.0006536$ adopted from \cite{fornari2020spatially}). Also, according to profile likelihoods on parameter distributions, each parameter was virtually identifiable. That is, the profile likelihood analysis of the fitted parameter ranges for the true parameter values of $r_a = 0.002413$, and $r_f = 0.0006536$, gave confidence intervals of $[0.02132-0.02476]$, and $[0.00538-0.00671]$, respectively. This demonstrates how the ABC technique with likelihood selects the appropriate underlying model structure from a list of options and determines the parameters with sufficient accuracy. In Fig. \ref{fig:9} (a), we provided the demo of the accepted parameters for the ADNI data for the strong astrocyte effect model ($sM2$) on birth-death as presented in Table \ref{tab:1}. In a similar fashion, we can estimate the parameters for the weak astrocyte effect model ($sM3$). For instance, as mentioned above, the ADNI data fit well per the mean, variance, and normal distributions, and the parameter range revealed that the accepted parameter values were close to the true values. However, due to the complexity of the ADNI data, there may be overfitting occurs when a model or simulation is too complex or has too many parameters relative to the available data. This scenario can be seen in Fig. \ref{fig:9} (b), where the parameter $r_a$ is converging initially but start to diverge as the number of iterations increases. Therefore, the number of iterations should be chosen wisely to balance the need for accuracy, stability, and convergence. In a similar fashion, using the Bayesian approach, we estimate the parameters for the weak astrocyte model ($sM3$) and simple birth-death model ($sM1$) presented in Table \ref{tab:1} using ADNI data. For the weak astrocyte effect model, the true parameter values of $r_a = 0.00531$, $r_f = 0.00004816$, gave confidence intervals of $[0.00491-0.00582]$, and $[0.00004173-0.00005100]$ and for a simple model, $r_a = 0.00231$, and $r_f = 0.00146$, gave confidence intervals of $[0.00199-0.00300]$, and $[0.00100-0.00211]$. Only the two unknown parameters ($r_a,r_f$) are shown in Fig. \ref{fig:9} (a) as our representative examples. The true parameter values for $A_{astro}$ for the present study are adopted from the previous knowledge \cite{whittington2018spatiotemporal,reid2020astrocytes,brandebura2022astrocyte,miller2018astrocyte,thuraisingham2022kinetic,carter2019astrocyte,fornari2020spatially,bedner2015astrocyte}. However, using the same knowledge, we supposed the values for $A_{astro}$ to predict the strong and weak astrocyte effects on $A\beta$ growth. 

\begin{figure}[htbp]
     \centering
     \begin{subfigure}[b]{0.48\textwidth}
         \includegraphics[width=\textwidth]{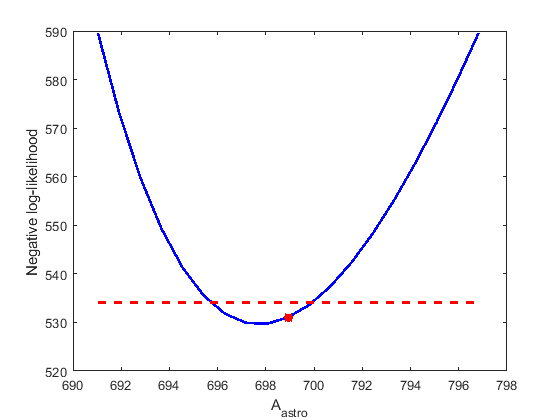}
         \caption{}
   \label{a}
     \end{subfigure}
     \begin{subfigure}[b]{0.48\textwidth}
         \includegraphics[width=\textwidth]{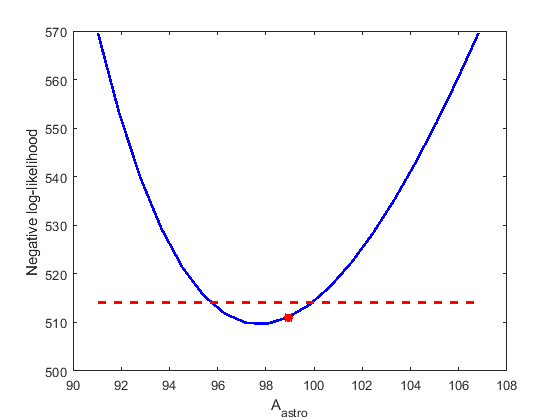}
         \caption{}
       \label{b}
     \end{subfigure}
         \caption{(Color online) The likelihood analysis for $A_{astro}$ for (a) strong astrocyte effect model ($sM2$), the red dot shows the parameter estimate of $A_{astro} = 698mL$ with the $95\%$ confidence intervals $[690–702]mL$ with true $A_{astro}=700\si{mL}$ and (b) weak astrocyte effect model ($sM3$), the red dot shows the parameter estimate of $A_{astro} = 98$ with the $95\%$ confidence intervals $[90–102]$ with true $A_{astro}=100\si{mL}$ as presented in Table \ref{tab:1} for the ADNI data. The red dotted line shows the maximum likelihood whereas, the y-axis represents the negative log-likelihood function value.}
        \label{fig:10}
\end{figure}
Next, for the strong and weak astrocyte effect models ($sM2-sM3$) to perform maximum likelihood parameter estimation for $A_{astro}$, we used the fminsearch function in MATLAB to minimize the negative log-likelihood as given in Eqs. (\ref{l2}). This method involves starting with an initial guess for the parameter values (in this case, $A_{astro}=700\si{mL}$ for $sM2$ and $A_{astro}=100\si{mL}$for $sM3$) and iteratively updating the parameter value until the true parameter value is achieved as shown in Fig. \ref{fig:10} (a-b). The estimated value of $A_{astro}$ for the strong astrocyte model in Fig. \ref{fig:10} (a) is very close to the true value, as indicated by the position of the red dot, which is considered a positive sign and suggests that the model is a good fit for the ADNI data, accurately capturing the underlying mechanisms that govern the relationship between astrocyte activation and brain function. Additionally, in a similar fashion, as depicted in Fig. \ref{fig:10} (b), for the weak astrocyte model, the estimated value (the red dot) of $A_{astro}$ is almost equal to the true value, it suggests that the model is a good fit for the ADNI data. The red dotted line shows the maximum likelihood as shown in Fig. \ref{fig:10} (a)-(b). It's important to note that while $A_{astro}$ may be a trivial function to minimize, the true value of $A_{astro}$ is not known a priori, and we estimated it from the data. In other words, the process of estimating $A_{astro}$ involves finding the parameter value that maximizes the likelihood of the observed data fitted in the model. In practice, this process can be challenging because there may be multiple parameter values that yield similar likelihood values, making it difficult to identify the true maximum likelihood estimate. Our study aims to examine the effects of weak and strong astrocytes, the differences in concentration ranges of astrocytes between Fig. \ref{fig:10} (a)-(b) are associated with the varying strengths of astrocytic effects. In this context, Fig. \ref{fig:10} (a) likely represents a scenario where astrocyte concentrations are higher, indicating a stronger presence of astrocytes. On the other hand, Fig. \ref{fig:10} (b) represents a situation with lower astrocyte concentrations, suggesting a weaker astrocytic presence. By correlating the concentration ranges of astrocytes in Fig. \ref{fig:10} (a)-(b) with the corresponding effects on $A\beta$ observed in our study, we explored how the strength of astrocytes influences the outcomes of interest.

Our data contained $1706$ patients, and we divided our job into many processors. We have selected the number of CPUs as the divisors of $1706$. As a result, for $10,000$ iterations, for example, for $1$ CPU, the computational time is $5012.1\si{s}$, for $2$ CPU's it is $140.3\si{s}$. To attain the required total time for each model, it often takes several million-time steps. If we solve the problem using conventional serial programming, this requires a significant amount of computing time. Through the use of open MPI and the C programming language, we can reduce computing time. We divide the sequential tasks involving the time step among available processors for each time iteration and perform them in parallel. All figures have been plotted and shown in Matlab after the data have undergone post-processing. To reduce the time required to acquire results for the parallel computation, we employed the SHARCNET supercomputer facilities (64 cores).

\section{Discussion}\label{dis}
To the best of our knowledge, this is the first study to analyze the effect of astrocytes on the $A\beta$ growth in AD patients using ADNI data. This paper discussed in detail the weak and strong effects of astrocytes on the increase of $A\beta$ utilizing ADNI data for AD patients per the $bl$ to $2$-year visits according to the ages. Additionally, we analyzed the growth of $A\beta$ and considered whether the astrocyte effect, as it is known from brain research, can be seen in deviations from exponential growth kinetics of $A\beta$ predicted by mathematical models that incorporate the astrocyte effect \cite{shaheen2021neuron,pal2022coupled,dossi2018human,urbanc1999dynamics,raj2015network}. Since $A\beta$ peptide can cause random oscillations in the growth kinetics of $A\beta$ (from one brain region to another) during the early stages of AD development, we constructed stochastic models that account for the astrocyte effect. Through the use of ADNI data for AD patients, we developed a framework for assessing the performance of stochastic models of $A\beta$ growth in AD. According to our results, we found that if the astrocyte effect is strong, that is the concentrations of astrocytes are greater than the concentrations of $A\beta$, then astrocytes help to clear the $A\beta$. Otherwise, the weak astrocyte effect, that is the low concentrations of astrocytes compared to $A\beta$, helps to promote the progression of AD. Importantly, the development of $A\beta$ deposits in AD patients is more likely a dynamic process that is rapidly expanding than a static phenomenon. Once significant $A\beta$ deposits have been created, they remain at a high level, if not treated.

Human astrocytes interact closely and dynamically with neurons, which enables them to regulate the homeostasis and functioning of the brain. Given their physiological roles, the involvement of astrocytes in the dynamics of neurodegenerative brain diseases seems logical \cite{dossi2018human}. Moreover, the interactions of astrocytes with neurons have been altered in a number of animal models of various neurological diseases \cite{bedner2015astrocyte,jo2014gaba,zhang2016purification}. This suggests that the dysfunction of astrocytes could play a role in the development or progression of these diseases. It is important to note that astrocytes are important for supporting and regulating the function of neurons, so any changes in their interactions could affect the overall functioning of the brain. Glial morphological and functional changes have been observed in human brain tissues from post-mortem or surgically resected patients who had a variety of brain disorders \cite{funato1998astrocytes,oberheim2006astrocytic}. We presented the strong and weak effects of astrocytes (i.e., $sM2$-$sM3$ models) in the human brains of AD patients using clinical data. However, choosing mean values of $A\beta$ as the initial conditions is a reasonable approach, but incorporating additional factors that contribute to AD progression could further refine the initial conditions and improve the accuracy of the simulation results. Using the ADNI dataset, we developed a fundamental framework for simulating the onset and progression of AD in a clinical trial to make it simpler to assess the efficacy of potential weak and strong effects of astrocytes. The release of inflammatory mediators that cause neurotoxicity, such as IL-1$\beta$, IL-6, and TNF-$\alpha$, was known to be increased by $A\beta$-induced astrocyte activation in the astrocyte-neuron co-culture, indicating a close connection between cytokine release and the neurotoxic events that take place at low initial levels of $A\beta$ \cite{garwood2011astrocytes}. Astrocytes and reactive astrocytes found near $A\beta$ plaques may be involved in local inflammation rather than direct $A\beta$ absorption and breakdown \cite{batarseh2016amyloid}. However, it is still unclear whether astrocytes truly migrate to the location of $A\beta$ plaques or if the shape of astrocytes already presents around the plaques simply changes. In recent research, it was shown that the distribution of astrocytes in the brains of AD animal models with $A\beta$ plaques was comparable to that of wild-type mice \cite{galea2015topological}. Since persistent activation of astrocytes in response to $A\beta$-associated inflammation is detrimental overall, reducing astrocyte activation may provide a novel therapeutic strategy to restore their supporting roles and halt additional inflammation-mediated cell death.

A Bayesian model is an effective tool for forecasting the onset and course of AD and for evaluating the effects of potential therapies in clinical trials, due to the complex and uncertain nature of the disease. Given that there are sufficient data to get accurate estimates of the model parameters, the model presented in this study should ideally be expanded to include more disease states. In addition, it is also known that gender, education, and genetic background (including the presence or absence of astrocytes) are all significant risk factors for AD. Additional elements, including family history, way of life, social standing, health, and environment, can also have an impact \cite{bilgel2019predicting}. So all these factors may also be beneficial to include in future studies. 

One additional important feature of the current research is that we demonstrated that the chance of identifying the true parameters and astrocyte effect may be challenging due to the high uncertainty in the clinical data. We note that the high level of uncertainty in the output of the simulations can be partially attributed to the sparsity and heterogeneity of the data that are currently available as well as the high variance in the measurements of the markers that have been used to categorize individuals into the various disease states of AD patients. This should potentially be taken into account as covariates in extended models because some of them can significantly affect the mean and variance of the model parameters that represent how $A\beta$ growth causes disease progression. The use of straightforward criteria to categorize people, such as how well they perform on cognitive tests and any possible measurement error that may be present, may affect the correct diagnosis and could play a significant role in the high variability \cite{bishara2023state}. Underlying medical conditions such as the presence of depression or psychosis, AD progression, inter-clinician differences in the application of diagnostic criteria, and resistance to cognitive decline due to cognitive reserve are some of the other factors that can introduce variability in measurements, both within and between individuals, and influence the diagnosis of the actual clinical stage \cite{hadjichrysanthou2018development, neumann2001measuring, abner2012mild, stern2003concept, schmand1997effects,oxtoby2018data}. Furthermore, it has been observed that detecting the genuine treatment impact may be difficult due to the significant uncertainty in clinical trial results, even if the treatment is highly effective \cite{mcdougall2021psychometric}. This might be one of the primary reasons for the failure of AD therapy trials.

Limitations of our models stem from the simplifying assumption of a single pathway of $A\beta$ changes from different states (i.e., NL to MCI) to AD using ADNI data. The longitudinal progression of AD is known to be heterogeneous. Clinical trials are challenging to run and expensive if large samples of patients are studied over a long period of time and if the initial concentrations of $A\beta$ are subject to change. Modelling and simulating AD trials are very low-cost activities that may present indispensable tools for refining the design of actual clinical trials and increasing the likelihood of correct treatment efficacy evaluations. Our results are consistent with the previous theoretical and experimental studies \cite{frost2017role,verkhratsky2010astrocytes,torok2022connectome,caberlotto2016integration}. It is also important to note that we defined our objective in this study as a description of the natural history of AD, which includes biological and cognitive changes triggered by astrocytes in addition to AD pathology. Using ADNI data, our modelling approach can estimate individual $A\beta$ growth trajectories and allows us to extract continuous information about the progression to AD. Individuals can be located along simulated trajectories using the developed framework provided by parameter estimation and Bayesian inference based on ADNI data. In addition, by obtaining estimates of astrocytes at the onset of AD, it may be possible to recruit participants more effectively for clinical trials targeting preclinical AD \cite{yiannopoulou2020current,song2022immunotherapy}. 
\section{Conclusions}\label{con}
Alzheimer's disease (AD) is probably the most common cause of dementia. Although the mechanism behind the onset and progression of AD is not fully known, mounting evidence points to a complex etiology that may involve a variety of genetic, environmental, and age-related variables. Increasing evidence suggests that astrocytes play an important and active role in many neurodegenerative disorders. Given the growing interest in normal astrocyte biology and the numerous research studies on variations in astrocyte function in the context of neurodegenerative diseases, it may come as a surprise that there are still no well-developed therapies that utilize astrocytes as primary targets.

One of two pathological hallmarks of AD is the deposition of amyloid-$\beta$ ($A\beta$) in the brain. The temporal evolution of $A\beta$ has been widely explored ex vivo. As a starting point, a logistic growth model can be used to simulate the in vivo temporal distribution of $A\beta$ in AD. Our ultimate objective was to concentrate on the stochastic growth model of $A\beta$ due to the intrinsic stochasticity of $A\beta$ concentrations. However, the structure of the functional forms of the kinetic equations describing the deposition of $A\beta$ proteins in the brain can be understood rather well within a deterministic framework. Therefore, firstly, we developed three deterministic models for $A\beta$ growth, viz., simple growth (Eq. \ref{eq1}), weak (Eq. \ref{astro}), and strong (Eq. \ref{astro1}) astrocyte effect models. Then, we applied a stochastic modelling approach to analyze the evolution of $A\beta$ in vivo using ADNI data from AD patients to investigate the effect of astrocytes on $A\beta$ spread in AD. We detailed the stochastic approach using the quantitative template for the progression of the AD project data set generated from the baseline to $2$-year samples visits from ADNI1 and ADNIGO/2. Specifically, we used the $A\beta$ data for $1706$ individuals with $6880$ visits and clinical follow-up visits to fall within the $24$-month window and analyzed the growth of $A\beta$ with respect to age. Importantly, for the present study, we chose the data for the patients who have already developed AD. The approximate Bayesian computation (ABC) technique was used to determine whether the aforementioned stochastic models (i.e., $sM1-sM3$ presented in Table \ref{tab:1} along with mean and variance presented in Table \ref{tab:2}) were fitted well or not. The stochastic trajectories of $A\beta$ data were sampled from the $bl$ to $2$-years, corresponding to the age utilized. 
In order to simulate the ADNI data as closely as possible, the growth of $A\beta$, as well as the mean and variance, were calculated at each time point. In summary, the effectiveness of astrocytes in clearing $A\beta$ in AD was studied using stochastic growth models fitted to ADNI data. The study found that a strong astrocyte effect, where astrocyte concentrations are greater than $A\beta$ concentrations, helps to clear $A\beta$, leading to a slower progression of AD. On the other hand, the weak astrocyte effect, where astrocyte concentrations are lower than $A\beta$ concentrations, promotes the progression of AD by enhancing $A\beta$ growth. The stochastic models used in the study were well-fitted and provided reliable parameter estimates. Therefore, it is clear that astrocytes play a crucial role in AD progression. The development of treatments that enhance strong astrocyte activation should be pursued to slow down AD progression. 

 Based on our observations, we have noticed a clear trend that as patients' age increases, the growth of $A\beta$ also increases. This increase in $A\beta$ growth has been observed in the absence of treatment. We compared and validated our results with the previous studies \cite{whittington2018spatiotemporal,frost2017role,sojkova2011longitudinal,villemagne2011longitudinal,vlassenko2011amyloid,verkhratsky2010astrocytes,torok2022connectome,caberlotto2016integration}. Integrating astrocytes as an essential criterion for AD medication developments can potentially result in more successful treatments for many millions of AD sufferers globally, given their crucial role in proper neuronal functioning. With the addition of more patients, such as MCI, EMCI, and NL, to clinical trial data, it should be possible to better predict the course of the disease and interpret model predictions more accurately in future studies. It will aid further in identifying the sources of variation, facilitating progress toward the more precise assessment of new preventative and therapeutic therapies for AD.

\section*{Acknowledgements}
The authors are grateful to the NSERC and the CRC Program for their support. RM is also acknowledging the support of the BERC 2022–2025 program and the Spanish Ministry of Science, Innovation and Universities through the Agencia Estatal de Investigacion (AEI) BCAM Severo Ochoa excellence accreditation SEV-2017–0718. This research was enabled in part by support provided by SHARCNET \url{(www. sharcnet.ca)} and Digital Research Alliance of Canada \url{(www.alliancecan.ca)}. The authors are grateful to Prof. Paul Marriott of the University of Waterloo for his insightful comments on the initial version of the manuscript.

Data collection and sharing for this project was funded by the Alzheimer's Disease Neuroimaging Initiative (ADNI) (National Institutes of Health Grant U01 AG024904) and DOD ADNI (Department of Defense award number W81XWH-12-2-0012). ADNI is funded by the National Institute on Aging, the National Institute of Biomedical Imaging and Bioengineering, and through generous contributions from the following: AbbVie, Alzheimer's Association; Alzheimer's Drug Discovery Foundation; Araclon Biotech; BioClinica, Inc.; Biogen; Bristol-Myers Squibb Company; CereSpir, Inc.; Cogstate; Eisai Inc.; Elan Pharmaceuticals, Inc.; Eli Lilly and Company; EuroImmun; F. Hoffmann-La Roche Ltd and its affiliated company Genentech, Inc.; Fujirebio; GE Healthcare; IXICO Ltd.; Janssen Alzheimer Immunotherapy Research $\&$ Development, LLC.; Johnson $\&$ Johnson Pharmaceutical Research $\&$ Development LLC.; Lumosity; Lundbeck; Merck $\&$ Co., Inc.; Meso Scale Diagnostics, LLC.; NeuroRx Research; Neurotrack Technologies; Novartis Pharmaceuticals Corporation; Pfizer Inc.; Piramal Imaging; Servier; Takeda Pharmaceutical Company; and Transition Therapeutics. The Canadian Institutes of Health Research is providing funds to support ADNI clinical sites in Canada. Private sector contributions are facilitated by the Foundation for the National Institutes of Health \url{(www.fnih.org)}. The grantee organization is the Northern California Institute for Research and Education, and the study is coordinated by the Alzheimer's Therapeutic Research Institute at the University of Southern California. ADNI data are disseminated by the Laboratory for Neuro Imaging at the University of Southern California.

\bibliographystyle{elsarticle-num}
\bibliography{references}

\end{document}